\pdfoutput=1
\PassOptionsToPackage{table,prologue}{xcolor}
\documentclass[letterpaper,journal]{IEEEtran}

\usepackage{times}
\usepackage{latexsym}
\usepackage{microtype}
\usepackage{inconsolata}
\usepackage{graphicx}
\usepackage[T1]{fontenc}
\usepackage{subfiles}
\usepackage{multirow}
\usepackage{siunitx}
\usepackage{adjustbox}
\usepackage{pgfplots}
\usepackage{booktabs} 
\usepackage{amsmath}
\usepackage{amssymb}
\usepackage{amsthm}
\usepackage[linesnumbered,ruled,vlined]{algorithm2e}
\SetKwBlock{Try}{try}{}
\SetKwBlock{Catch}{except}{}
\usepgfplotslibrary{groupplots}
\usetikzlibrary{calc}
\usepackage{xcolor}
\usepackage[utf8]{inputenc}  
\usepackage{tikz}
\usepackage{pgfplots}

\usepackage[most]{tcolorbox}
\usepackage{subcaption}
\usepackage{listings}
\definecolor{codegreen}{rgb}{0,0.6,0}
\definecolor{codegray}{rgb}{0.5,0.5,0.5}
\definecolor{codepurple}{rgb}{0.58,0,0.82}
\definecolor{backcolour}{rgb}{0.95,0.95,0.92}
\definecolor{academic-blue}{RGB}{31,119,180}
\definecolor{academic-red}{RGB}{214,39,40}
\definecolor{academic-green}{RGB}{44,160,44}
\definecolor{academic-yellow}{RGB}{255,204,0}
\definecolor{academic-orange}{RGB}{255,127,14}
\definecolor{academic-purple}{RGB}{148,103,189}

\lstdefinestyle{mystyle}{
    language=Python,
    aboveskip=3mm,
    showstringspaces=false,
    columns=flexible,
    numbers=none,
    backgroundcolor=\color{backcolour},
    commentstyle=\color{codegreen},
    keywordstyle=\color{magenta},
    numberstyle=\tiny\color{codegray},
    stringstyle=\color{codepurple},
    basicstyle=\small\ttfamily,
    breakatwhitespace=false,         
    breaklines=false,                 
    captionpos=b,                    
    keepspaces=false,                 
    numbersep=5pt,                  
    showspaces=false,                
    showstringspaces=false,
    showtabs=false,                  
    tabsize=2,
    escapeinside=``
}
\definecolor{LightGray}{gray}{0.9}
\definecolor{Green}{rgb}{0,0.6,0}
\definecolor{LightPink}{rgb}{1.0,0.9,0.9}
\definecolor{Red}{rgb}{0.8,0,0}
\definecolor{Green}{rgb}{0,0.6,0}
\definecolor{codegray}{gray}{0.9}
\definecolor{keywordcolor}{rgb}{0.13,0.13,1.0}
\definecolor{stringcolor}{rgb}{0.58,0.0,0.82}
\definecolor{desiredcolor}{rgb}{1,0,0}  
\definecolor{assertcolor}{rgb}{0.8,0,0}
\definecolor{LightCyan}{rgb}{0.88,1,1}
\pgfplotsset{compat=1.18}
\graphicspath{ {./figures/} }
\newcommand{\tool}{\texttt{VALTEST}}

\usepackage{pifont}

\lstdefinelanguage{Python}{
    keywords={class, def, return, if, else, elif, for, while, in, is, not, and, or, try, except, finally, raise, import, from, as, with, del, global, nonlocal, lambda, yield, assert},
    keywordstyle=\color{blue}\bfseries,
    identifierstyle=\color{black},
    sensitive=true,
    comment=[l]{\#},
    commentstyle=\color{gray}\ttfamily,
    string=[b]",
    stringstyle=\color{red}\ttfamily,
    morecomment=[s]{"""}{"""},
    morecomment=[s]{'''}{'''},
    literate={True}{{{\color{cyan}True}}}1
             {False}{{{\color{cyan}False}}}1
             {None}{{{\color{cyan}None}}}1,
    morekeywords={pd,DataFrame,unittest,TestCase,setUp,self,assert,output,task_func} 
}

\definecolor{yellowhighlight}{rgb}{1,1,0.8}

\usepackage{hyperref}
\usepackage{amsmath,amsfonts}
\usepackage{array}
\usepackage{textcomp}
\usepackage{stfloats}
\usepackage{url}
\usepackage{verbatim}
\usepackage{graphicx}
\usepackage{cite}

\begin{document}
\title{Toward Automated Validation of Language Model Synthesized Test Cases using Semantic Entropy}

\author{%
\IEEEauthorblockN{Hamed Taherkhani, Jiho Shin, Muhammad Ammar Tahir, Md Rakib Hossain Misu, Vineet Sunil Gattani, Hadi Hemmati}
\thanks{
 Hamed Taherkhani and Jiho Shin are with the Lassonde School of Engineering, York University, Toronto, Ontario, Canada. Emails: hamedth@yorku.ca, jihoshin@yorku.ca
}
\thanks{Muhammad Ammar Tahir is with the University of Washington, Seattle, USA. Email: ammartahircs@gmail.com}
\thanks{Md Rakib Hossain Misu is with the University of California, Irvine, USA. Email: mdrh@uci.edu}
\thanks{Vineet Sunil Gattani is with Arizona State University, USA. Email: vgattan1@asu.edu}
\thanks{Hadi Hemmati is an associate professor at Lassonde School of Engineering, York University, Toronto, Ontario, Canada. Email: hemmati@yorku.ca}
}




\maketitle

\begin{abstract}
Modern Large Language Model (LLM)-based programming agents often rely on test execution feedback to refine their generated code. These tests are synthetically generated by LLMs. However, LLMs may produce invalid or hallucinated test cases, which can mislead feedback loops and degrade the performance of agents in refining and improving code. This paper introduces {\tool}, a novel framework that leverages \textbf{semantic entropy} to automatically \textbf{validate test cases} generated by LLMs. 
By analyzing the semantic structure of test cases and computing entropy-based uncertainty measures, {\tool} trains a machine learning model to classify test cases as valid or invalid and filters out invalid test cases. Experiments on multiple benchmark datasets and various LLMs show that {\tool} not only boosts test validity by up to 29\% but also improves code generation performance, as evidenced by significant increases in pass@1 scores. Our extensive experiments also reveal that semantic entropy is a reliable indicator to distinguish between valid and invalid test cases and provides a robust solution for improving the correctness of LLM-generated test cases used in software testing and code generation.
\end{abstract}


\section{Introduction}
\label{Introo}
\begin{figure*}[ht]
    \centering
    \begin{minipage}{\textwidth}
    \begin{lstlisting}[basicstyle=\footnotesize,showstringspaces=false]
def max_area(height):
""" You are given an integer array height of length n. There are n vertical lines drawn such that the two endpoints of the ith line are (i, 0) and (i, height[i]). Find two lines that together with the x-axis form a container, such that the container contains the most water. Return _the maximum amount of water a container can store_.**Notice** that you may not slant the container.
**Example 1:**Input:** height =[1,8,6,2,5,4,8,3,7]**Output:** 49
**Explanation:** The above vertical lines are represented by array [1,8,6,2,5,4,8,3,7]. In this case, the max area of water (blue section) the container can contain is 49. 
**Example 2:**Input:** height = [1,1]**Output:** 1"""
    \end{lstlisting}
    \end{minipage}
    
    \raggedright 

    \footnotesize Tests Generated by GPT4o
    \hspace{1cm}
    \begin{tikzpicture}
        \draw[->, thick] (0,0) -- (0,-0.3); 
    \end{tikzpicture}
    
    \begin{tikzpicture}

    \node (box2) [draw, rectangle, inner sep=1pt, align=left, text width=0.49\textwidth, anchor=north west] at (0,0) {
    \begin{minipage}[t]{\linewidth}
    \begin{lstlisting}[basicstyle=\footnotesize]
                      Function Input   Expected Output
1. assert max_area ([1,3,2,5,25,24,5])   ==   24 (*@\cmark@*)
2. assert max_area([1,2,3,4,5,6,7,8,9,1])==   8 (*@\xmark@*)
3. assert max_area([1,2,3,4,5,6,7,1,8,9])==   18 (*@\xmark@*)
4. assert max_area([1,1,1,1,1,1,1,1,1,1])==   9 (*@\cmark@*)
    \end{lstlisting}
    \end{minipage}
    };

    \node (box3) [draw, rectangle, inner sep=1pt, align=left, text width=0.45\textwidth, anchor=north west] at ($(box2.north east)+(0.8cm,0)$) {
    \begin{minipage}[t]{\linewidth}
    \begin{lstlisting}[basicstyle=\footnotesize]
 Token Entropy of the Input & Output
1. [1.9, 0.1, 0, 0.2, 0.1, 0, 0], [(*@\textcolor{Green}{0.1}@*)]
2. [2, 2, 1.3, 1, 0.5, 1.7, 1, 0.5, 1.5, 2], [(*@\textcolor{Red}{1.3}@*)]
3. [2, 2, 1.5, 1.3, 1, 0.9, 0.9, 1.2, 0.7, 0], [(*@\textcolor{Red}{2}@*)]
4. [0.6, 0.4, 0.4, 0.1, 0, 0, 0, 0, 0.1, 0.5], [(*@\textcolor{Green}{0}@*)]
    \end{lstlisting}
    \end{minipage}
    };

    \draw[->, thick] ($(box2.east)+(0, 0.15cm)$) -- ($(box3.west)+(0, 0.15cm)$);

    \end{tikzpicture}

    \caption{An example of test case generation using GPT4o. The check mark indicates a valid test and the cross mark indicates an invalid test. The entropy of function input and expected output parts are displayed on the right.}
    \label{fig:1}
\end{figure*}

LLMs are actively employed in various software development tasks, including software testing, design, requirements engineering, code generation, maintenance, deployment, and many more \cite{10449667,10440574}. One important task is automated code generation. Recent studies have explored multi-agent and collaborative frameworks such as Reflexion~\cite{shinn2024reflexion}, LATS~\cite{zhou2023language}, AgentCoder~\cite{huang2023agentcoder}, EPiC~\cite{taherkhani2024epic}, and LDB~\cite{zhong2024ldb} to enhance code generation. These agents generate feedback from test execution on the generated code and try to refine the code based on the generated feedback. They iteratively refine the code by leveraging feedback from test execution, often generating their own test cases for internal validation while reserving dataset-provided tests for final evaluation. The underlying assumption is that tests are unavailable during internal evaluation, necessitating synthetic test generation. While this approach is conceptually sound, it introduces a significant risk: \textbf{invalid test cases} (e.g., tests with incorrect assertions or logic) may produce misleading feedback. For example, Figure \ref{fig:1} represents a sample of test cases generated by GPT-4o using a function description in natural language. In this example, the LLM produced two valid and two invalid assertion statements (while more tests can be generated, we present only four for brevity). The correct outputs for the second and third assertions should be 20 (not 8), and 25 (not 18), respectively.

An invalid test case will provide incorrect feedback to the agent, potentially leading valid code implementations toward invalid ones or further misdirecting invalid implementations, making them deviate even more from the correct solution. Invalid or hallucinated tests pose two major challenges in LLM-based code generation: (1) false positives, where defective code passes erroneous tests, and (2) false negatives, where correct code fails due to flawed tests \cite{shinn2024reflexion}. As demonstrated in Section \ref{validity-perf}, the validity of tests has a significant impact on code generation performance. Despite this, existing studies have largely overlooked the critical issue of test validation in code generation workflows. In this paper, we address this critical question: \textbf{How can we determine whether an LLM-generated test case is valid when the code under test is unavailable or its correctness is unknown?}

To address this challenge, we use a hallucination detection approach to predict invalid test cases generated by LLMs. Hallucination in the context of natural language generation refers to the phenomenon where models generate text that is either nonsensical or unfaithful to the provided source content~\cite{ji2023survey}. Many recent works employ entropy as a potential method for LLM hallucination detection~\cite{kadavath2022language, ledger2024detecting, quevedo2024detecting, huang2023look, fadeeva2024fact, varshney2023stitch, orgad2024llms, farquhar2024detecting, kossen2024semantic}.

Entropy is a probabilistic measure of uncertainty. Semantic entropy has been shown to be an effective approach to detect hallucinations~\cite{farquhar2024detecting} in LLMs. This method quantifies uncertainty based on the semantic content of the generated text rather than the entire text. Semantic entropy is a probabilistic approach that is computed over the meanings of the text rather than the sequence of words. A high semantic entropy score suggests that the model produces semantically diverse and uncertain responses, which may indicate hallucinations. We adopt this approach to detect hallucination and leverage it to validate LLM-generated tests. In Figure \ref{fig:1}, the token entropy corresponding to the function's input and the expected output of each test case in the box on the left is presented in the box on the right. For instance, in the test case \lstinline!assert maxarea([1, 3, 2, 5, 25, 24, 5]) == 24!, the token entropy of the expected output token \textit{24} is $0.1$, while the entropy values for the function input tokens $1, 3, 2, 5, 25, 24, 5$ are $1.9, 0.1, 0, 0.2, 0.1, 0, 0$, respectively. Comparing the token entropy scores of tests 1 and 4 with those of tests 2 and 3, we observe that invalid tests exhibit higher entropy values, either in the function input or the expected output tokens. This is expected, as LLMs are prone to generating invalid test cases when they are uncertain about the assertions' input or output, often as a result of hallucination. This observation motivated us to develop {\tool}, a tool that leverages entropy to validate LLM-generated tests. {\tool} supports Python class-based unit tests in addition to single line assertions.

We show the effectiveness of {\tool} in improving code generation by (1) evaluating {\tool} in identifying invalid tests and (2) incorporating {\tool} into LLM-based code generation agents.

To evaluate its ability to identify invalid tests, we first generated multiple test suites using different LLMs and datasets.
After generating the tests, we identified the semantic components of each test case and computed their corresponding entropy values. Subsequently, we derived multiple feature sets that capture the statistical properties of these entropy values. We executed each test case on its ground truth code (the given solution in the dataset) to label each case as either valid or invalid. We trained a k-fold ensemble model to predict test validity using labeled features, discarded tests predicted as invalid, and evaluated the resulting test suites using validity rate, mutation score, and code coverage before and after filtering tests.

To show the impact of {\tool} on improving code generation, we integrated our test validation process into two code generation frameworks (Reflexion~\cite{shinn2024reflexion} and Language Agent Tree Search (LATS)~\cite{zhou2023language}) as case studies to assess the extent to which validation of generated test cases will improve performance. For this assessment, we used a recent code generation dataset, \textbf{BigCodeBench} \cite{zhuo2024bigcodebench} on top of a SOTA reasoning model (OpenAI's o3-mini~\footnote{\href{https://openai.com/index/openai-o3-mini/}{https://openai.com/index/openai-o3-mini/}}).

Our results show that {\tool} improves the validity rate of test suites from $8\%$ to $29\%$ and also boosts \textit{pass@1}~\cite{chen2021evaluatinglargelanguagemodels} on Reflexion by $7\%$ and $11\%$, depending on the dataset. Our results also show that semantic entropy consistently outperforms all the baselines in Section \ref{4.2}.

The main contributions of this paper include:
\begin{enumerate} 
    \item To the best of our knowledge, {\tool} is the first study to investigate the problem of predicting the validity of test cases generated by LLMs.
    \item We demonstrate the effectiveness of {\tool} in test case validation across common benchmark datasets and SOTA LLMs.
    \item We demonstrate the effectiveness of {\tool} in improving code generation performance by incorporating {\tool} into an agentic code generation tool.
\end{enumerate}
We also released the data and source code for our experiments to facilitate replication and extension by other researchers (\href{https://github.com/HamedTaherkhani/VALTEST}{https://github.com/HamedTaherkhani/VALTEST}).

\section{Related Work}
\label{relatedWork}
\subsection{LLM-based Test Case Generation}
Traditional test case generation approaches utilize search-based \cite{harman2009theoretical, delgado2022interevo}, constraint-based \cite{xiao2013characteristic}, or random-based \cite{pacheco2007feedback} strategies to maximize code coverage. However, these methods are often sub-optimal in terms of creating maintainable test cases. With the advent of LLMs, recent studies have explored using LLMs to generate more human-readable unit test cases by learning from developer-written tests in their large training sets. Most of these studies focus on improving the effectiveness of generated test cases, e.g., by pre-training or fine-tuning on code-related tasks \cite{tufano2020unit, alagarsamy2024a3test, hashtroudi2023automated, rao2023cat}, leveraging reinforcement learning \cite{steenhoek2023reinforcement}, designing effective prompts \cite{xie2023chatunitest, dakhel2024effective, zhang2023well,daghighfarsoodeh2025deep}, incorporating documentation \cite{plein2024automatic, vikram2023can}, and integrating with search-based methods \cite{lemieux2023codamosa}. Despite initial successes, research indicates that LLM-generated test cases face challenges such as correctness issues \cite{yuan2023no, guilherme2023initial, li2023prompting}, and low coverage on certain benchmarks, highlighting the need for continued refinement and comparison with traditional methods like EvoSuite \cite{tang2024chatgpt, bhatia2024unit} and Pynguin \cite{lemieux2023codamosa, bhatia2024unit}. More recent works provide notable examples of the advancements in automated test case generation. Liu et al. introduced AID~\cite{liu2024llm}, combining LLMs with differential testing to generate fault-revealing test cases for programs that have already passed traditional test suites. Alagarsamy et al.'s A3Test~\cite{alagarsamy2024a3test} utilizes domain adaptation to produce accurate, assertion-informed test cases, emphasizing test name consistency and verification. ChatTester~\cite{yuan2023no} refines unit tests generated by ChatGPT through iterative feedback loops to improve test accuracy. Meanwhile, Ouédraogo et al.~\cite{ouedraogo2024large} compare various LLMs with EvoSuite, assessing test generation in terms of coverage, readability, and correctness.

\subsection{Refining LLM-generated Test Cases}
Several studies in the literature address generating valid test cases and refining them. They rely on the ground truth to check the test case validity. Guilherme and Vincenzi ~\cite{guilherme2023initial} conducted an initial investigation on test case generation using OpenAI's LLM for Java programs. They compared the generated test cases with traditional tools like EvoSuite. They noted that the LLM performed comparably well, but highlighted the importance of refinement to improve fault detection.
Li et al.~\cite{li2023prompting} focused on generating unit tests by prompting a GPT-4-based LLM using the QuixBugs dataset. They observed that iterative prompting was key to improving the correctness. The model often generated valid test cases but needed an iterative refinement process to handle compilation errors.
Yuan et al.~\cite{yuan2023no} evaluated the unit test generation capabilities of ChatGPT and introduced ChatTester, an approach that refines test cases iteratively. They found that while ChatGPT could generate tests with high readability, the tests often suffered from correctness issues.
Li et al., 2024~\cite{li2024large} proposed TestChain, which decouples the generation of test inputs and outputs. This allows for a multi-agent system, where one agent generates test inputs, and another computes the expected outputs through a Python interpreter, thus improving the correctness of the outputs. Sollenberger et al. \cite{sollenberger2024llm4vv} introduced LLM4VV, which exploits LLMs as judges to evaluate compiler test suites for parallel programming models like OpenMP and OpenACC. They introduced intentional errors (negative probing) to assess the ability to detect issues in code. CODET~\cite{chen2022codetcodegenerationgenerated} is a test-driven approach that selects the most correct code solution among the candidates generated by a language model. It first prompts the same model to generate test cases and then evaluates each code solution by executing it against these tests. Using a dual execution agreement strategy, CODET identifies consensus sets and ranks them based on the number of agreeing solutions and tests passed. These studies use code execution feedback from the correct source code to generate valid test cases, but in the code generation task, this is not given (it's the target solution). 
In contrast, {\tool} uses an entropy-based approach to predict the validity of generated tests without the need for the code under test or the test execution. Predicting the validity of generated tests is a task that has not been explored in the existing literature.

\subsection{Hallucination Detection}
There are several metrics to detect hallucinations in LLMs, e.g., statistical metrics, information extraction-based metrics, natural language inference-based metrics, question-answering-based metrics, and model-based metrics~\cite{ji2023survey}. 
Zhou et al.~\cite{zhou2020detecting} proposed a method to detect hallucinations at the token level, aiming to predict whether each token in generated outputs is hallucinated or faithful to the source input. They create synthetic hallucinated data using a pre-trained model (BART) to fine-tune models for hallucination detection. Raunak et al. 
\cite{raunak2021curious} examined two main types of hallucinations: those triggered by perturbations to the input and those caused by noise in the training data. By linking these hallucinations to the long-tail theory of learning, the authors showed that samples more deeply memorized by the model are more likely to produce hallucinations when perturbed. 
Hallucinations in code generation by LLMs are addressed in several papers.
De-Hallucinator~\cite{eghbali2024hallucinator} tackled the problem by iteratively refining prompts with relevant API references drawn from the model's initial output, reducing errors in API usage and improving accuracy.
Rahman et al.~\cite{rahman2024code} introduced HallTrigger, a technique that deliberately triggers arbitrary code hallucination in LLMs to analyze their frequency and impact.
HALLUCODE~\cite{liu2024exploring} took a broader approach, creating a taxonomy of hallucinations across five categories and developing a benchmark to assess code LLMs' effectiveness in recognizing and mitigating hallucinations. CodeHalu~\cite{tian2024codehalu} built on this by employing execution-based verification to detect hallucinations in generated code, classifying them into mapping, naming, resource, and logic errors.
In this paper, we use semantic entropy to detect hallucinations. Unlike prior code generation studies that rely on task-specific hallucination definitions and mitigation strategies, our approach uses a common method to identify hallucinations. To our knowledge, {\tool} is the first to use hallucination detection to identify invalid test cases.
\begin{figure*}
    \centering
    \includegraphics[width=\linewidth]{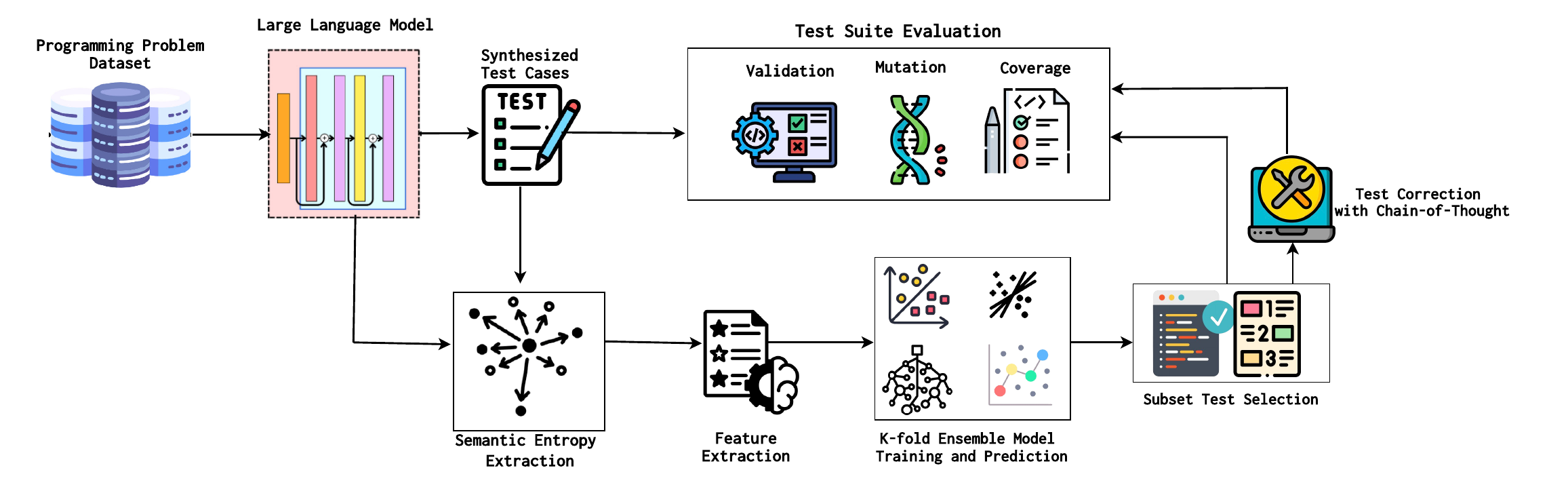}
    \caption{Overall approach of {\tool}}
    \label{overall_fig}
\end{figure*}
\section{VALTEST}

{\tool} systematically validates LLM-generated tests and enhances test case generation for LLM-based systems (see Figure \ref{overall_fig}). {\tool} begins by generating tests, along with the probability assigned to each token produced by the LLM. These tests are dynamically labeled against the code based on execution results. {\tool} finds the semantics of the tests and uses statistical entropy measures at the level of meaning to train a predictive model, which filters likely invalid tests and retains valid ones. The model is trained and evaluated on the labeled dataset (valid/invalid) using a k-fold ensemble method, where each time, it considers k-1 folds as the train set and the remaining fold as the evaluation set. We classify a test as hallucinated or invalid if its output probability falls below a predefined \textit{threshold} applied to the model’s predictions.
The rest of this section provides a detailed step-by-step explanation of {\tool}.


\subsection{Test Case Generation and Processing}
For each function $f$ with signature $\mathit{sig}(f)$ and docstring $\mathit{doc}(f)$, {\tool} generates test cases:
\begin{equation}
    \mathcal{T}_f = \{t_1, ..., t_n\} \quad \text{where } 10 \leq n \leq 20.
\end{equation}
Different functions vary in terms of difficulty, and thus, they may require a different number of tests to fully validate the function. We set both a minimum (10) and a maximum (20) limit for the number of test cases and allow the LLM to decide how many test cases are necessary. 
Each test $t_i$ is associated with the token sequence 
$\mathbf{w}_i = (w_{i1}, \dots, w_{im})$ and the top five token 
candidates with their probabilities, denoted as 
$P(w_{ij}) = \{(w_{ij}^n, p_{ij}^n) \}_{n=1}^{5}$.
We obtain these token candidates and probabilities using the 
\texttt{transformers} library or OpenAI’s API. After generating the test cases, we parse them into an AST (abstract syntax tree) representation to discard syntactically incorrect tests.
\begin{equation}
    \mathcal{T}'_f = \{ t_i \in \mathcal{T}_f \mid \textsc{AST-Parse}(t_i) \neq \emptyset \}.
\end{equation}

At the end of this step, we have a test suite $\mathcal{T}'_f$ where every function has multiple tests with probability information, and all tests are free of syntax errors.

\subsection{Initial Test Suite Evaluation}
In this step, we evaluate the generated test suite in terms of validity, mutation score, and code coverage (see Section ~\ref{metric}). First, we run each test on its corresponding function's ground-truth code (the solution given in the dataset) to annotate each test with valid or invalid labels and measure the validity rate for the test suite. We consider a test case valid if it runs against the ground-truth code without producing any errors. In addition, we measure the mutation score for the test suite as an evaluation metric, along with code coverage. We use the Mutmut\footnote{\href{https://github.com/boxed/mutmut}{https://github.com/boxed/mutmut}} testing tool for mutation testing. Mutmut supports various types of mutations, such as operator mutation, number mutation, string mutation, lambda mutation, keyword mutation, etc. 
In this research, we used the default setting that includes all types of mutations.

\subsection{Semantic Entropy Extraction and Feature Calculation}

Test semantics are the parts of the test case that specify the function's input arguments or the expected output of the test. For each test $t_i$, we identify and extract its semantics using Algorithm~\ref{alg:semantic-extraction-try-catch}. In this algorithm, we first find the assertion statement in the test case and then extract both the function's input argument $\mathbf{a}_i^{\text{in}}$ and the expected output $\mathbf{a}_i^{\text{out}}$:
\begin{equation}
    \mathbf{a}_i^{\text{in}},\mathbf{a}_i^{\text{out}} = test\_semantic\_extraction(t_i).
\end{equation}

For instance, in \lstinline!assert max_area([1,3,2,5,25,24,5])==27!, we identify ``[1,3,2,5,25,24,5]'' as the function input argument and ``27'' as the expected output. Our approach supports both \textbf{Python class-based unit tests and simple assertions}.

Once the function input and expected output are identified, we proceed to match these strings with tokens in a predefined list of (token, probability) pairs ($\mathbf{w}_i$).
To achieve this, we employ a greedy string-matching algorithm that attempts to match substrings from one sequence with tokens from another sequence. This approach is widely used in lexical analysis and tokenization processes~\cite{alfred2007compilers}. We use $greedy\_token\_matching$ in Algorithm~\ref{matching} to extract $\mathbf{W}_i^{\text{in}}$ and $\mathbf{W}_i^{\text{out}}$:
\begin{equation}
\resizebox{0.95\linewidth}{!}{$
    \mathbf{W}_i^{\text{in}}, \mathbf{W}_i^{\text{out}} = greedy\_token\_matching(\mathbf{a}_i^{\text{in}},\mathbf{a}_i^{\text{out}}, \mathbf{w}_i).$}
\end{equation}

Each token \(w_{ij}\) includes the top five candidates and the probability scores associated with them, which will enable us to calculate the token entropy. During the matching process, we ignore punctuation tokens. These tokens are considered noise as they do not contribute to the assertion outcomes, and they shouldn't be considered as the semantics of the test. We compute Shannon's entropy for each matched token 
\(w_{ij}\) using:
\begin{equation}
    H(w_{ij}) = -\sum_{n=1}^{5} p_{ij}^n \log p_{ij}^n,
\end{equation}
which quantifies the uncertainty of the model’s token prediction. To capture the overall semantic uncertainty of the test case, we summarize $\mathbf{W}_i^{\text{in}}$ and $\mathbf{W}_i^{\text{out}}$ using five statistical measures: the mean (\(\mu\)), maximum (\(\max\)), minimum (\(\min\)), total (sum, \(\Sigma\)), and variance (\(\sigma^2\)). This results in the following feature vector for test \(t_i\):
\begin{equation}
\small 
\begin{split}
\mathbf{x}_i &= \big[\underbrace{\mu(H^{in}), \max(H^{in}), \min(H^{in}), \Sigma (H^{in}), \sigma^2(H^{in})}_{\text{Input Entropy Features}},\\
&\underbrace{\mu(H^{out}), \max(H^{out}), \min(H^{out}), \Sigma (H^{out}), \sigma^2(H^{out})}_{\text{Output Entropy Features}}\big],
\end{split}
\end{equation}
where, $H^{in} = H(\mathbf{W}_i^{\rm in})$ and $H^{out} = H(\mathbf{W}_i^{\rm out})$ for test $t_i$.



\subsection{K-fold Training and Classification}
Next, we employ a k-fold cross-validation method to train and validate our test cases. In this method, the dataset is split into \( k \) folds $\{\mathcal{F}_1, ..., \mathcal{F}_k\}$. The model is trained on \( k-1 \) folds and evaluated on the remaining fold. The split is function-based, meaning that for each validation iteration, the test cases corresponding to a subset of functions (\( k-1 \) folds) are selected for training, while the remaining fold is used for evaluation.  
We use a soft voting ensemble model, which is composed of several models, including \textit{logistic regression}, \textit{support vector machine}, \textit{random forest}, \textit{XGBoost}, \textit{LightGBM}, \textit{AdaBoost}, and \textit{gradient boosting}. We empirically compared the performance of the individual classifiers during our preliminary experiments and observed that the ensemble consistently outperformed the individual classifiers. Based on these observations, we chose the ensemble approach for our framework.

After training, the ensemble model is used to predict outcomes on the validation set. Using the k-fold validation approach, the model predicts whether each test case is valid or not. The model generates a score between 0 and 1 for each test case, and a \textit{threshold} is set to determine validity. If the score exceeds this \textit{threshold}, the test case is considered valid. We observed that some functions in the dataset may not have any test cases passing the \textit{threshold}. To address this issue, we employed the \textit{topN} parameter to select the top \( N \) test cases for functions where fewer than \( N \) test cases pass the \textit{threshold}. The \textit{topN} parameter is essential as it allows us to select a non-empty subset of test cases for each function. Without test cases, a function's mutation score and coverage would be zero, leading to an undesirable outcome.

\subsection{Test Case Subset Selection}
Finally, we apply the previously mentioned \textit{topN} and \textit{threshold} ($\theta$) parameters to select a subset of tests from the initial set:
\begin{equation}
\mathcal{T}_f^{\text{valid}} = \{t_i \in \mathcal{T}'_f \mid \hat{p}_i \ge \theta\}.
\end{equation}
Then, the final set is given by

\begin{equation}
\small
    \mathcal{T}_f^{\text{final}} = \mathcal{T}_f^{\text{valid}} \cup 
\begin{cases}
\displaystyle \underset{\mathcal{S} \subseteq \mathcal{T}'_f \setminus \mathcal{T}_f^{\text{valid}}}{\text{argmax}} \; \sum_{t_i \in \mathcal{S}} \hat{p}_i, & \text{if } |\mathcal{T}_f^{\text{valid}}| < N, \\[2ex]
\varnothing, & \text{if } |\mathcal{T}_f^{\text{valid}}| \ge N,
\end{cases}
\end{equation}

with the additional constraint that 
\begin{equation}
|\mathcal{S}| = N - |\mathcal{T}_f^{\text{valid}}|.
\end{equation}
This subset will undergo a final evaluation based on three key metrics: validity rate, mutation score, and code coverage.

\section{Experimental Setup}
\label{experiments}
\subsection{Datasets}
We use five datasets in this study: \textbf{HumanEval}~\cite{chen2021evaluatinglargelanguagemodels} (164 programming problems with function signatures, docstrings, solutions, and unit tests), \textbf{MBPP-sanitized}~\cite{austin2021program} (427 curated Python tasks for novices, with problem statements, solutions, and 3 tests each),
\textbf{LeetCode} (512 interview-style coding challenges\footnote{\href{https://leetcode.com/problemset/}{https://leetcode.com/problemset/}}), \textbf{TestEval}~\cite{wang2025testevalbenchmarkinglargelanguage}, and \textbf{BigCodeBench}~\cite{zhuo2024bigcodebench}.

HumanEval, MBPP, and LeetCode were selected for their popularity and importance. BigCodeBench is a more recent and more complex dataset. TestEval is a dataset focused on test generation and is specifically analyzed for data leakage. 

TestEval uses GPT-4o as a case study and compare two subsets of problems: (i) 21 problems released after the model’s training cut-off, and (ii) 21 older problems released before the cut-off. For the post–cut-off problems, they report that none of the official test cases appear in the LLM-generated tests, whereas for the pre–cut-off subset, some official tests are found among the generated ones. However, since the evaluation produces 20 tests per problem, they argue that the fraction of copied tests is small, and thus the leakage effect is minimal. Our use of TestEval follows the same core evaluation that underlies their leakage analysis: we also generate up to 20 tests per problem. Coincidentally, we use the same models evaluated in TestEval (GPT-4o, GPT-3.5-turbo, Llama 3.1, and CodeQwen). Therefore, any residual contamination could only affect a very small subset of the generated tests. Consistent with TestEval’s own findings, {\tool} is not meaningfully impacted by possible contamination in this benchmark.


\subsection{Models}
We evaluated four LLMs varying in size and open/closed source models. OpenAI's gpt-3.5-turbo-0125 and gpt-4o-2024-08-06, Meta's Llama-3.1-8B-Instruct, and Alibaba's Qwen2.5-Coder-32B-Instruct.
Additionally, we use OpenAI's o1-preview, a reasoning-focused model for invalid test case analysis in Section \ref{app:causal_analysis} and OpenAI's o3-mini for Sections \ref{validity-perf} and \ref{RQ2}.

\subsection{Evaluation Metrics}
\label{metric}
To evaluate the proposed approach, we employ two categories of metrics: one assessing the validity of the generated test cases and another evaluating test case adequacy. For the former, we introduce \textbf{validity rate (VR)} to measure the test suite's validity. For the test adequacy metrics, we employ two widely recognized classic test adequacy metrics: \textbf{Mutation Score (MS)} and Average \textbf{Line Coverage (LC)}, which are also commonly used to evaluate LLM-generated test cases \cite{shin2024assessing}. In addition, we use \textbf{\textit{pass@1}} metric~\cite{chen2021evaluatinglargelanguagemodels} in Section~\ref{RQ2}.

\subsubsection{Validity Rate}
Let \( F = \{ f_1, f_2, \dots, f_n \} \) represent the set of functions, where \( n \) is the total number of functions. For each function \( f_i \), let \( T_i = \{ t_{i1}, t_{i2}, \dots, t_{im_i} \} \) represent the set of \( m_i \) test cases designed to test that function.

A test case is considered \textbf{valid} if it executes successfully against the ground-truth implementation without producing any errors (e.g., assertion failures or runtime exceptions). Conversely, a test case is \textbf{invalid} if it fails during execution due to incorrect assertions, logical errors, or hallucinated outputs that do not align with the expected behavior of the function.

We define a binary function \( V(t_{ij}) \), where:
\begin{equation}
V(t_{ij}) =
  \begin{cases}
   1 & \text{if the test case } t_{ij} \text{ is valid for function } f_i, \\
   0 & \text{otherwise.}
  \end{cases}
\end{equation}
We define the \textit{Validity Rate} (VR) across all functions as the rate of valid test cases to the total number of test cases for all functions. Formally, this is expressed as:
\begin{equation}
{\text{VR}} = \frac{\sum_{i=1}^{n} \sum_{j=1}^{m_i} V(t_{ij})}{\sum_{i=1}^{n} m_i}.
\end{equation}

\subsubsection{Precision and Recall}
\label{pre_recall}
We formally define Precision and Recall in the context of {\tool}'s task. {\tool} acts as a binary classifier where the goal is to distinguish between ``valid'' and ``invalid'' test cases. We define the positive class as a test case being \textbf{valid}.

The classification outcomes are as follows:
\begin{itemize}
    \item \textbf{True Positives (TP):} Tests that are actually valid and are correctly classified as valid by {\tool} (i.e., they are kept).
    \item \textbf{False Positives (FP):} Tests that are actually invalid but are incorrectly classified as valid by {\tool} (i.e., they are kept).
    \item \textbf{True Negatives (TN):} Tests that are actually invalid and are correctly classified as invalid by {\tool} (i.e., they are discarded).
    \item \textbf{False Negatives (FN):} Tests that are actually valid but are incorrectly classified as invalid by {\tool} (i.e., they are discarded).
\end{itemize}

\noindent \textbf{Precision} measures the purity of the tests that are kept by the filter. It is the fraction of tests classified as valid that are actually valid.
\begin{equation}
    \text{Precision} = \frac{\text{TP}}{\text{TP} + \text{FP}}.
    \label{eq:precision}
\end{equation}

\noindent \textbf{Recall} measures the completeness of the filtering process. It is the fraction of all truly valid tests that are successfully identified and kept.
\begin{equation}
    \text{Recall} = \frac{\text{TP}}{\text{TP} + \text{FN}}.
    \label{eq:recall}
\end{equation}

In the context of our work, \textbf{Precision is functionally equivalent to our primary metric, Validity Rate (VR)}. The numerator of the precision formula (TP) represents the number of valid tests in the final, filtered test suite. The denominator (TP + FP) represents the total number of tests in that same filtered suite. This ratio is precisely the definition of Validity Rate. Therefore, all Validity Rate (VR) results presented throughout our experiments can be directly interpreted as the precision of the {\tool} filtering mechanism. Recall, conversely, quantifies the trade-off, measuring the proportion of good tests that are preserved.

\subsubsection{Mutation Score (MS)}
Mutation testing is performed using the Mutmut mutation testing tool, which generates mutants through various operators, including operator mutation and number mutation, among others. Mutmut generates mutants for each function, and the MS is calculated as the percentage of killed mutants relative to the total mutants generated across all functions.

For each function \( f_i \), a set of mutants \( M_i \) is generated, with \( |M_i| \) representing the total number of mutants for function \( f_i \). A \emph{killed mutant} refers to a modified version of the original function that is detected and ``killed'' by the test suite, meaning the test cases can identify the injected fault and produce a failing result. Out of these \( |M_i| \) mutants, \( K_i \) represents the number of mutants that were killed by the test suite.

The mutation score \( MS \) across all functions is calculated as the ratio of the total number of killed mutants to the total number of generated mutants. Mathematically, this can be expressed as:
\begin{equation}
MS = \frac{\sum_{i=1}^{n} K_i}{\sum_{i=1}^{n} |M_i|}.
\end{equation}

\subsubsection{Average Line Coverage (LC)}
We assess line coverage by executing test cases on each function, calculating line coverage for each function individually, and then averaging these values across all functions to report as the LC.
Let \( L(f_i) \) represent the set of lines of code in function \( f_i \), and let \( L(t_{ij}) \subseteq L(f_i) \) represent the subset of lines covered by test case \( t_{ij} \).

The set of lines covered by all test cases for a function \( f_i \) is given by the union of the lines covered by each test case:

\begin{equation}
C(f_i) = \bigcup_{j=1}^{m_i} L(t_{ij}).
\end{equation}

The \textit{coverage ratio} for a function \( f_i \) is then defined as the ratio of the number of distinct lines covered by the test cases to the total number of lines in the function:

\begin{equation}
\text{Coverage}(f_i) = \frac{|C(f_i)|}{|L(f_i)|}.
\end{equation}

The average line coverage (LC) across all functions is defined as the mean of the coverage ratios for all functions:

\begin{equation}
\text{LC} = \frac{1}{n} \sum_{i=1}^{n} \text{Coverage}(f_i).
\end{equation}

\subsubsection{Pass@k}
We evaluated Reflexion using the \textit{pass@k} metric from \cite{chen2021evaluatinglargelanguagemodels}. Unlike CodeBLEU and ROUGE, which focus on textual semantics, \textit{pass@k} measures code functionality. In Equation~\ref{eq:1}, \(\mathbb{E}\) is the expected value over problems, \(n\) the total generated samples per task, and \(c\) the number of correct samples. It estimates the probability that at least one of the top \(k\) samples is correct per problem, averaged across all problems.

\begin{equation}
pass@k := \underset{\text{Problems}}{\mathbb{E}}\left[1 - \frac{\binom{n-c}{k}}{\binom{n}{k}}\right].
\label{eq:1}
\end{equation}

\subsubsection{Mann-Whitney U test}
The Mann-Whitney U test was used to compare the distributions of two independent samples. This non-parametric test evaluates differences in central tendencies by analyzing rank distributions. The p-value in this test indicates the likelihood of observing a test statistic as extreme as the calculated \textit{U} value if the null hypothesis (no difference in distributions) holds. A small p-value (typically $<$ 0.05) suggests that the rank differences are unlikely due to chance, supporting a statistically significant difference between the groups.

\subsection{Wilcoxon signed-rank test}
To assess the statistical significance of performance differences between {\tool} (Semantic Entropy) and baseline methods, we employ the \textit{Wilcoxon signed-rank test}. The Wilcoxon signed-rank test is a non-parametric statistical hypothesis test used to compare paired samples, serving as a robust alternative to the paired $t$-test when normality assumptions cannot be guaranteed. Unlike parametric tests, it does not assume that the data follow a normal distribution; instead, it assumes only that the distribution of differences between paired observations is symmetric. In our evaluation, we apply this test to compare the VR of Semantic Entropy against each baseline method across all 16 experiments (4 datasets $\times$ 4 LLMs), treating each dataset-LLM combination as a paired observation. We formulate the null hypothesis ($H_0$) as equal distributions between methods, and the alternative hypothesis ($H_1$) as Semantic Entropy achieving superior performance (one-sided test, $\alpha = 0.05$). The test computes a test statistic by ranking the absolute differences between paired observations and summing the ranks of positive differences, with smaller $p$-values indicating stronger evidence against the null hypothesis.

\subsection{Computational Overhead}
VALTEST introduces virtually no latency into an LLM-based workflow.  
\textbf{Feature extraction} is a single linear pass over each test, computing ten entropy statistics. These statistics come directly from the token probabilities returned during test synthesis, so \textbf{no additional model calls are required}.  
Training the $k$-fold ensemble of classical learners operates on a tiny feature matrix and finishes within minutes on a standard laptop-class CPU. At inference time, the model performs one vector lookup per test, making inference \textbf{effectively instantaneous}.

\subsection{Research Questions}
We define the following research questions (RQ) to systematically evaluate the effectiveness of {\tool}:

\begin{itemize}
    \item \textbf{RQ1:} How effective is VALTEST in validating LLM-generated test cases?
    \item \textbf{RQ2:} To what extent does the test suite’s validity rate impact code generation performance?
    \item \textbf{RQ3:} How effective is VALTEST in improving the performance of LLM-based code generation agents?
    \item \textbf{RQ4:} How does VALTEST compare with baseline approaches for filtering invalid test cases?
    \item \textbf{RQ5:} How does the selection of \textit{threshold} and \textit{topN} hyperparameters impact the balance between MS and VR?
    \item \textbf{RQ6:} {What is the impact of different feature sets in {\tool}?}
    \item \textbf{RQ7:} {How to combine {\tool} and Chain-of-Thought (CoT) reasoning to improve both validity rate and mutation score of test suites?}
\end{itemize}

\subsection{Experiment Design}
\label{design}
\textbf{RQ1:} \textbf{For the first experiment}, we use the \textit{topN} and \textit{threshold} parameters to filter a subset of test cases from the initial set, selecting those that are more likely to be valid based on the predictions made by our model. We generate a total of 16 test suites (4 datasets * 4 LLMs). We evaluate {\tool} using two approaches: \textbf{cross-dataset validation} and \textbf{k-fold cross-validation}. In the cross-dataset approach, three datasets are used for training, while a third distinct dataset is reserved for evaluation. For example, when evaluating the HumanEval-GPT-4o test set, we train on LeetCode-GPT-4o, MBPP-GPT-4o, and TestEval-GPT-4o, ensuring that the model remains blind to HumanEval-GPT-4o during training. The goal of this approach is to assess whether {\tool} generalizes well and is transferable across different datasets. In k-fold cross-validation, we use the same dataset and perform training on \(k - 1\) folds while evaluating on the remaining fold. We report VR, LC, and MS metrics to assess the effectiveness of {\tool}. We use a \textit{threshold} of $0.75$ and a \textit{topN} of $3$ as hyper-parameters.

\textbf{RQ2 and RQ3: }We adapted two popular code generation agents, \textbf{Reflexion}~\cite{shinn2024reflexion} and LATS (Language Agent Tree Search)~\cite{zhou2024languageagenttreesearch}. Reflexion processes test feedback through verbal reasoning and then maintains the feedback in a dynamic episodic memory buffer for subsequent trials. Unlike Reflexion, which employs an iterative, linear self-correction loop, LATS explores multiple reasoning paths through a tree-search algorithm, representing a distinct and more complex class of agentic problem-solving. They both use test generation agents to generate tests. The BigCodeBench dataset is designed to work with Python’s unittest framework. To ensure Reflexion’s and LATS's compatibility with this dataset, we modified both agents to use unittest instead of simple assertions. This also makes our integration process standard and extensible to any agent using the unittest framework. 

\textbf{In RQ2}, we examine how the validity rate of the test suite impacts the performance of Reflexion. We construct five test suites of fixed size, each corresponding to a different validity rate. To mitigate the effects of randomness in LLMs, five independent test suites are generated per validity level. These suites are integrated into Reflexion to evaluate \textit{pass@1}, with results averaged across the five runs. This experimental setup allows for a robust assessment of whether increased test suite validity correlates with improved code generation performance.

\textbf{In RQ3}, we evaluate {\tool}'s effectiveness in improving \textbf{code generation}. In this experiment, we first evaluate the adapted Reflexion and LATS using unit tests without any validation mechanism. Next, we re-evaluate both agents using test cases validated and filtered by {\tool} to see how much our test validation approach improves code generation compared to the previous experiment. 

In both \textbf{RQ2 and RQ3}, we use GPT-4o for test case generation and OpenAI’s recent reasoning model, o3-mini, for code generation and refinement. Both experiments were conducted on the BigCodeBench dataset. We selected the \textbf{BigCodeBench} dataset due to two main reasons: (1) older benchmarks such as HumanEval, MBPP, and LeetCode exhibit high baseline \textit{pass@1} scores, limiting the potential for measurable improvements; and (2) \textbf{BigCodeBench} is a newer dataset with significantly lower baseline \textit{pass@1}, making it more suitable to assess the impact of {\tool} on code generation.

\textbf{In RQ4}, we compare {\tool} (Semantic Entropy) with five baselines in terms of (1) Validity Rate (Precision), (2) Recall, (3) Mutation Score and Line Coverage, and (4) pass@1 score on a downstream code generation task. The baselines are: \textbf{Naive Entropy}, \textbf{Semantic Probability}, \textbf{FirstN}, \textbf{Basic Semantic Entropy}, and \textbf{Chain-of-Thought (CoT)}. We set the \textit{threshold} to $0.75$ and the \textit{topN} to $3$ for this RQ. We use Reflexion agent and BigCodeBench dataset for the downstream code generation comparison with the baselines.


\textbf{Computational Cost of Chain-of-Thought Baseline.} CoT incurs substantial computational overhead compared to {\tool} and other baselines. Specifically, CoT requires a separate LLM call for each test case to perform step-by-step reasoning and validation. In our experiments on TestEval with GPT-4o, the base test suite contained 3,404 test cases. At an estimated 1,000 additional tokens per test case for CoT validation, this translates to approximately 3,400,000 additional tokens consumed for validation. This amounts to roughly \$30 in additional costs for a single dataset. In contrast, {\tool} (and other baselines) require no additional LLM calls and complete validation in milliseconds on standard hardware.

\begin{table*}[ht]
\centering
\caption{Comparison of number of test cases (Tests), the percentage of identified valid test (\%Tests), Validity Rate (VR), Line Coverage (LC), and Mutation Scores (MS), across datasets and LLMs under Base and VALTEST settings. Average values across LLMs are highlighted for each dataset, and percentage increases in validity rates are indicated in green. VR and VR-X represent the validity rate using k-fold cross-validation and cross-dataset validation respectively.}
\label{table:1}
\begin{tabular}{
    ll
    S[table-format=4.0]
    l
    S[table-format=1.3]
    S[table-format=1.2]
    S[table-format=4.0]
    S[table-format=1.2] 
    l
    l
    l
    S[table-format=1.3]
    l
}
\toprule
\multirow{2}{*}{Dataset} & \multirow{2}{*}{LLM} & \multicolumn{4}{c}{Base} & \multicolumn{7}{c}{VALTEST} \\
\cmidrule(lr){3-6} \cmidrule(lr){7-11}
& & {\#Tests} & {VR} & {LC} & {MS} & {\#Tests} & {\%Tests} & {VR (Precision)} & {Recall} & {VR-X} & {LC} & {MS} \\
\midrule
HE & GPT-4o & 3157 & 0.83 & 0.96  & 0.87  & 2512 & 0.8 & 0.92{\color{Green}(+$9$\%)} & 0.88 & 0.94 & 0.96 & 0.86{\color{Red}(-$1$\%)} \\
HE & GPT-3.5 Turbo & 2688 & 0.74 & 0.95 & 0.83 & 1649 & 0.56 & 0.90{\color{Green}(+$16$\%)} & 0.74 & 0.92 & 0.94 & 0.79{\color{Red}(-$4$\%)} \\
HE & Llama 3.1 8B & 2459 & 0.63 & 0.93 & 0.80 & 645 & 0.30 & 0.77{\color{Green}(+$14$\%)}& 0.32 &0.78 &0.87 & 0.63{\color{Red}(-$17$\%)} \\
HE & Qwen2.5-Coder & 3199 & 0.84 & 0.98 & 0.88  & 2466 & 0.77 & 0.94{\color{Green}(+$10$\%)} & 0.86 & 0.97& 0.98 & 0.86{\color{Red}(-$2$\%)} \\
\rowcolor{LightCyan}
HE & \textbf{Average} & \textbf{2875} & \textbf{0.76} & \textbf{0.95} & \textbf{0.84} & \textbf{1818} & \textbf{0.63} & \textbf{0.88}{\color{Green}(+$12$\%)} & \textbf{0.70} &\textbf{0.90} &\textbf{0.94} & \textbf{0.78}{\color{Red}(-$6$\%)} \\
\midrule
LeetCode & GPT-4o & 9047 & 0.75 & 0.98 & 0.86  & 5651 & 0.62 & 0.94{\color{Green}(+$19$\%)} & 0.78 & 0.95& 0.98 & 0.85{\color{Red}(-$1$\%)} \\
LeetCode & GPT-3.5 Turbo & 8214 & 0.63 & 0.98 & 0.83 & 3430 & 0.42 & 0.90{\color{Green}(+$27$\%)}& 0.60 &0.90 & 0.97 & 0.82{\color{Red}(-$1$\%)} \\
LeetCode & Llama 3.1 8B & 5023 & 0.46 & 0.95 & 0.78 & 1058 & 0.21 & 0.75{\color{Green}(+$29$\%)} & 0.34 & 0.72& 0.89 & 0.72{\color{Red}(-$6$\%)} \\
LeetCode & Qwen2.5-Coder & 9084 & 0.66 & 0.98 & 0.84 & 4643 & 0.51 & 0.91{\color{Green}(+$25$\%)} & 0.70 & 0.93 &0.98 & 0.82{\color{Red}(-$2$\%)}\\
\rowcolor{LightCyan}
LeetCode & \textbf{Average} & \textbf{7842} & \textbf{0.62} & \textbf{0.97} & \textbf{0.83} & \textbf{3695} & \textbf{0.47} & \textbf{0.87}{\color{Green}(+$25$\%)} & \textbf{0.61} &\textbf{0.87} & \textbf{0.95} & \textbf{0.80}{\color{Red}(-$3$\%)} \\
\midrule
MBPP & GPT-4o & 7750 & 0.70 & 0.85 & 0.82 & 3581 & 0.46 & 0.83{\color{Green}(+$13$\%)} & 0.55 &0.78 & 0.81 & 0.78{\color{Red}(-$4$\%)} \\
MBPP & GPT-3.5 Turbo & 5934 & 0.60 & 0.82 & 0.78 & 1991 & 0.34 & 0.74{\color{Green}(+$14$\%)} & 0.41 & 0.73 & 0.73 & 0.71{\color{Red}(-$7$\%)} \\
MBPP & Llama 3.1 8B & 4461 & 0.53 & 0.77 & 0.72 & 1202 & 0.27 & 0.61{\color{Green}(+$8$\%)} & 0.31 &0.61 & 0.67 & 0.61{\color{Red}(-$11$\%)} \\
MBPP & Qwen2.5-Coder & 7010 & 0.67 & 0.87 & 0.83 & 1892 & 0.27 & 0.82{\color{Green}(+$15$\%)} & 0.33 &0.78 & 0.83 & 0.77{\color{Red}(-$6$\%)} \\
\rowcolor{LightCyan}
MBPP & \textbf{Average} & \textbf{6288} & \textbf{0.62} & \textbf{0.83} & \textbf{0.79} & \textbf{2166} & \textbf{0.34} & \textbf{0.75}{\color{Green}(+$13$\%)} & \textbf{0.40} & \textbf{0.73} & \textbf{0.76} & \textbf{0.72}{\color{Red}(-$7$\%)} \\
\bottomrule

TestEval & GPT-4o & 3404 & 0.57 & 0.92 & 0.73 & 1468 & 0.43 & 0.74{\color{Green}(+$17$\%)} & 0.56 & 0.75 & 0.90 & 0.69{\color{Red}(-$4$\%)} \\
TestEval & GPT-3.5 Turbo & 2499 & 0.39 & 0.77 & 0.63 & 839 & 0.33 & 0.61{\color{Green}(+$22$\%)} & 0.51 & 0.61 & 0.75 & 0.59{\color{Red}(-$4$\%)}\\
TestEval & Llama 3.1 8B & 1706 & 0.38 & 0.65 & 0.59 & 511 & 0.30 & 0.58{\color{Green}(+$20$\%)} & 0.38 & 0.57 & 0.63 & 0.53{\color{Red}(-$6$\%)}\\
TestEval & Qwen2.5-Coder & 1460 & 0.75 & 0.88 & 0.72 & 1044 & 0.71 & 0.86{\color{Green}(+$11$\%)} & 0.81 & 0.86 & 0.88 & 0.71{\color{Red}(-$1$\%)}\\
\rowcolor{LightCyan}
TestEval & \textbf{Average} & \textbf{2267} & \textbf{0.52} & \textbf{0.81} & \textbf{0.67} & \textbf{966} & \textbf{0.45} & \textbf{0.70}{\color{Green}(+$18$\%)} & \textbf{0.57} & \textbf{0.70} & \textbf{0.79} & \textbf{0.63}{\color{Red}(-$4$\%)} \\

\bottomrule
\end{tabular}

\end{table*}

\textbf{RQ5: }In general, there is a trade-off between the validity rate and the mutation score as the \textit{threshold} and \textit{topN} hyper-parameters are adjusted. This trade-off is the focus of \textbf{RQ5}, where we select one test suite (GPT-3.5-turbo on HumanEval) and modify the hyperparameters to identify potentially optimal values. We do not exhaustively experiment on all test suites, as the goal is not to find the best configuration across all datasets and LLMs, but rather to explore the trade-off and to show the potential impact of these hyperparameters. Please note that there are several hyper-parameter optimization methods that exist in the literature \cite{zamani2019revisiting, zamani2020cost,zamani2021pragmatic}, and a proper study of these methods is not within the scope of this paper. We use \textit{threshold} values of 0.5, 0.65, 0.8, and 0.85, and \textit{topN} values of 1, 3, 5, and 7 for this ablation study.

\textbf{RQ6: }In this \textbf{RQ}, we investigate the impact of distinct feature sets on model performance. Two feature sets are employed, representing (1) the function input arguments in test cases and (2) the expected outputs, abbreviated as \textit{FI} (\textit{Function Input}) and \textit{EO} (\textit{Expected Output}), respectively. To evaluate distributional differences between valid and invalid test cases, we apply the Mann-Whitney~U test to the \textit{FI} and \textit{EO} feature sets. This statistical test determines whether significant differences exist between valid and invalid test cases with respect to these metrics. For each feature set, we compute a representative feature defined as the mean of entropy values, denoted as \textit{function\_input\_mean} (\textit{FI\_mean}) and \textit{expected\_output\_mean} (\textit{EO\_mean}). All analyses were conducted using Python’s \textit{SciPy} library. Finally, to assess the individual contributions of these feature sets to \tool's effectiveness, we train the model under three configurations: (1) \textit{FI} alone, (2) \textit{EO} alone, and (3) all features combined. The evaluation follows the experimental setup of prior sections, utilizing the same 16 test suites derived from 4 LLMs and 4 benchmark datasets.

\textbf{RQ7: }In the \textbf{seventh experiment}, rather than discarding invalid test cases, we applied a CoT reasoning approach to correct them step by step. In this approach, we prompt the LLM with a step-by-step example demonstrating how to reason about a test case and, if necessary, correct the test case output. This method allows us to correct the invalid test cases and add them to the set of predicted valid test cases. We then measure the validity rate, mutation score, and code coverage again, and compare these results with those from the main results in Section \ref{RQ1}. For this evaluation, we use GPT-4o as it showed the best results and has better reasoning abilities than the other models, making it the most suitable LLM for this experiment.

\section{Results}

\subsection{\textbf{RQ1: How effective is {\tool} in validating LLM-generated test cases?}}
\label{RQ1}
The results for this experiment are presented in Table \ref{table:1}, divided into two columns: ``Base'' and ``{\tool}''. The column ``Base'' represents the initial set of generated test cases without filtering, while the column ``{\tool}'' represents the test suite after applying {\tool}, where a subset of test cases is selected from the initial set. We compare the number of tests, VR, LC, and MS between the Base and {\tool} test suites. VR and VR-X represent the validity rate using k-fold cross-validation and cross-dataset validation, respectively. The number of tests in the {\tool} suite is smaller than in the Base suite, as {\tool} filters and selects a subset of test cases that are more likely to be valid. As a result, VR increases across all experiments, while MS and LC decrease. This trade-off between increased VR and decreased LC and MS depends on the selected \textit{threshold} and \textit{topN} hyper-parameters. The decrease in LC and MS is due to the smaller number of test cases in the {\tool} suite compared to the Base suite. 

\begin{figure*}[t!]
  \footnotesize
  \centering
  \captionsetup[subfigure]{labelformat=empty}
    \begin{subfigure}[b]{0.45\textwidth}
    \centering
    \begin{tikzpicture}
      \begin{axis}[
        width=\columnwidth,
        height=0.6\columnwidth,
        title={BigCodeBench-hard},
        xlabel={Validity (\%)},
        ylabel={Pass@1},
        xmin=55, xmax=105,
        ymin=0.32, ymax=0.45,
        xtick={60,70,80,90,100},
        ytick={0.33,0.35,0.37,0.39,0.41,0.43},
        grid=both,
        major grid style={line width=0.2pt,draw=gray!50},
        minor grid style={line width=0.1pt,draw=gray!20},
        axis background/.style={fill=white},
        tick align=outside,
      ]
        \addplot[
          color=blue,
          mark=square*,
          line width=1.5pt,
          mark size=3pt,
        ] coordinates {
          (60,0.366)
          (70,0.361)
          (80,0.36092)
          (90,0.4013)
          (100,0.407)
        };
      \end{axis}
    \end{tikzpicture}
  \end{subfigure}
  \hfill
    \begin{subfigure}[b]{0.45\textwidth}
    \centering
    \begin{tikzpicture}
      \begin{axis}[
        width=\columnwidth,
        height=0.6\columnwidth,
        title={BigCodeBench-full},
        xlabel={Validity (\%)},
        ylabel={Pass@1},
        xmin=55, xmax=105,
        ymin=0.55, ymax=0.72,
        xtick={60,70,80,90,100},
        ytick={0.56,0.58,0.60,0.62,0.64,0.66,0.68,0.70},
        grid=both,
        major grid style={line width=0.2pt,draw=gray!50},
        minor grid style={line width=0.1pt,draw=gray!20},
        axis background/.style={fill=white},
        tick align=outside,
      ]
        \addplot[
          color=blue,
          mark=square*,
          line width=1.5pt,
          mark size=3pt,
        ] coordinates {
          (60,0.5878)
          (70,0.6081)
          (80,0.6311)
          (90,0.6554)
          (100,0.6865)
        };
      \end{axis}
    \end{tikzpicture}
  \end{subfigure}
  \captionsetup{belowskip=0pt}
  \caption{Impact of test suite validity rate on code generation performance in Reflexion for BigCodeBench-hard and BigCodeBench-full sets. Pass@1 is the average pass@1 for executions over 5 random test subsets of size 500 with a specific VR.}
  \label{fig:combined}
\end{figure*}

The average increase in VR for the HumanEval, LeetCode, MBPP, and TestEval datasets is 12\%, 25\%, 13\%, and 18\% respectively.

On the HumanEval dataset, the average MS decrease is 6\%, with a decrease of only 1\% for GPT-4o, 4\% for GPT-3.5-turbo, and 2\% for QwenCoder. However, for Llama 3.1, the MS decrease is 17\%, indicating that the \textit{threshold} value of 0.75 may not be optimal for this test suite. The number of selected test cases for this suite in {\tool} is 645, which is only 30\% of the original test cases. For the Llama 3.1 model on HumanEval, a higher \textit{threshold} could result in more balanced outcomes. The trade-off between VR and MS is further analyzed in Section \ref{app:threshNtopn} while keeping all hyperparameters consistent for this experiment. In summary, {\tool} shows a considerable improvement without compromising the quality of test cases, especially if a proper \textit{threshold} is selected. On the LeetCode dataset, the average MS decrease is only 3\%. The decrease for GPT-4o and GPT-3.5-turbo is just 1\%, while for Llama 3.1, it is 6\%. This is due to the ratio of selected test cases for Llama 3.1 being only 21\%, compared to 62\% and 42\% for GPT-4o and GPT-3.5-turbo, respectively. This suggests that, similar to the Llama 3.1 test suite for HumanEval, using a lower \textit{threshold} on the Llama 3.1 test cases for the LeetCode dataset could lead to a better trade-off between MS and VR. Overall, the results on LeetCode demonstrate a substantial improvement in test case validity, with only a slight decrease in mutation score. The average MS decrease is 7\% on MBPP, while the VR increase is 13\%, demonstrating a substantial improvement in validity consistent with the trends observed on HumanEval and LeetCode. Our results on TestEval also demonstrate substantial improvements in validity rate across all models. For GPT-4o, the validity rate increased from 57\% to 74\%, representing a 17 percentage improvement. Similarly, with GPT-3.5-turbo, VALTEST improved the validity rate by 22 points, and with Llama-3.1-8B, by 20 points.

\textbf{VR vs VR-X.} Comparing VR-X and VR values, we observe that the performance of {\tool} on cross-dataset evaluation is comparable to that of k-fold cross-validation. Specifically, cross-dataset evaluation yields a 2\% improvement in HumanEval and a 2\% decrease in MBPP relative to k-fold cross-validation. These results suggest that {\tool} is transferable across datasets without training on the target dataset, and generally without significant performance degradation.

\textbf{Precision and Recall.} While {\tool} achieves high precision, indicating that most of the tests it retains are indeed valid, the recall values are comparatively lower, showing that some valid tests are discarded. For example, on the HumanEval benchmark, the average precision is 88\% whereas the average recall is 70\%. Notably, precision is equivalent to the validity rate (VR), as both measure the proportion of retained tests that are truly valid. Despite the modest recall, this does not compromise the overall quality of the test suite. This is supported by the consistently high MS and LC observed across datasets. These metrics confirm that even with fewer retained tests, the resulting test suites remain highly adequate. Therefore, very high recall is not strictly necessary when the objective is to preserve test adequacy.

\textbf{In conclusion, {\tool} significantly improves the validity of the test suite without compromising its effectiveness, and it can be transferred to other datasets without training on the target dataset.}


\subsection{\textbf{RQ2: To what extent does the test suite's validity rate impact code generation performance?}}
\label{validity-perf}

We investigated the impact of test validity on a code generation agent called Reflexion. For our evaluation, we used the \textit{pass@1} metric in the two subsets of the BigCodeBench dataset: Full and Hard. We synthesized multiple test suites with different validity rates of 60\%, 70\%, 80\%, 90\%, and 100\% and ran the adapted Reflexion using each separately. For each VR, we sampled 5 test subsets of size 500 from the initial Base test suite (Table \ref{table:1}).  We measured the average \textit{pass@1} of the five test subsets and reported in Figure \ref{fig:combined} for both full and hard sets in BigCodeBench. There is a clear positive correlation between the validity of test cases and \textbf{code generation performance (\textit{pass@1})}. Reflexion performs significantly better when given a test suite with higher validity, confirming that \textbf{higher test validity directly translates to improved code generation outcomes}.

In Figure~\ref{fig:combined}, the pass@1 score for BigCodeBench-hard exhibits non-monotonic behavior, with a slight decrease when the validity rate increases from 60\% to 80\%. This pattern likely stems from limitations in how Reflexion processes feedback from test suites on this particularly challenging dataset. BigCodeBench-hard requires complex reasoning about intricate function behaviors, and in such scenarios, even a small number of invalid tests can significantly mislead Reflexion's iterative refinement process. The pass@1 score begins to improve consistently only after a threshold validity rate is achieved (above 80\%), where the remaining invalid tests become too sparse to significantly disrupt the refinement process. This observation reinforces the importance of achieving high validity rates, particularly for challenging benchmarks, to ensure that agentic code generation systems receive reliable feedback signals.

\textbf{Realistic Fault Detection Through Reflexion.} It is important to note that the code generation process in Reflexion implicitly evaluates realistic fault detection. During each iteration of Reflexion's debugging loop, the agent generates multiple incorrect code implementations before converging to a correct solution. These incorrectly generated code snippets represent realistic faults produced by the LLM. The test cases must effectively distinguish between these flawed implementations and the correct solution to provide useful feedback for iterative improvement. Our results demonstrate that test suites with higher validity rates lead to significantly improved pass@1 scores (see Figure~\ref{fig:combined}). This improvement indicates that valid tests are more effective at detecting the realistic faults present in LLM-generated incorrect code during the debugging process.

\begin{table*}[ht]
\centering
\caption{Comparison of the number of test cases (\#Tests), Validity Rate (VR), Line Coverage (LC), and Mutation Score (MS) on BigCodeBench under both Base and {\tool} settings. Pass@1 scores represent the performance of Base tests with Reflexion and {\tool}'s tests with Reflexion.}
\label{table:100}
\setlength{\tabcolsep}{4pt}
\begin{tabular}{
    l
    S[table-format=4.0]
    l
    S[table-format=1.2]
    S[table-format=1.2]
    S[table-format=1.2]
    S[table-format=1.2]
    S[table-format=4.0]
    S[table-format=1.2]
    l
    l
    S[table-format=1.3]
    l
    l
}
\toprule
\multirow{2}{*}{Dataset} & \multicolumn{6}{c}{Base} & \multicolumn{7}{c}{VALTEST} \\
\cmidrule(lr){2-7} \cmidrule(lr){8-14}
& {\#Tests} & {VR} & {LC} & {MS} & {Pass@1(Ref)} & {Pass@1(LATS)} & {\#Tests} & {\%Tests} & {VR} & {LC} & {MS} & {Pass@1(Ref)} & {Pass@1(LATS)} \\
\midrule
BCBHard& 1250 & 0.63 & 0.83 & 0.55 & 0.30 & 0.32 & 512 & 0.41 & 0.77{\color{Green}(+$14$\%)} & 0.77 & 0.47 & 0.37 {\color{Green}(+$7$\%)} & 0.40 {\color{Green}(+$8$\%)}\\
BCB & 1591 & 0.64 & 0.94 & 0.58 & 0.56 & 0.60 & 624 & 0.39 & 0.85{\color{Green}(+$21$\%)} & 0.91 & 0.50 & 0.67{\color{Green}(+$11$\%)} & 0.72{\color{Green}(+$12$\%)}\\
\bottomrule
\end{tabular}

\end{table*}
\begin{figure*}[ht]
  \centering
  \begin{tikzpicture}[every label/.style={font=\small}]
    \begin{groupplot}[
      group style={
        group size=4 by 4,
        horizontal sep=1.2cm,
        vertical sep=1.2cm,
        group name=myplots,
      },
      width=4.2cm,
      height=4cm,
      ybar,
      xtick=\empty,
      ymin=40, ymax=100,
      nodes near coords,
      nodes near coords style={font=\footnotesize},
      major tick length=0pt,
      enlarge x limits={abs=0.6cm},
      axis x line=bottom,
      axis y line=left,
      xlabel style={font=\small, yshift=-5pt},
      every axis plot/.append style={
        bar width=9pt,
        line width=0.4pt,
        draw=black!30,
      },
      legend style={
        draw=none,
        fill=none,
        legend columns=3,
        font=\footnotesize,
        cells={anchor=west},
        /tikz/every even column/.append style={column sep=8pt},
        at={($(myplots c1r1.north)+(0,15pt)$)},
        anchor=south,
      },
      legend to name=commonlegend,
    ]
    \footnotesize

    \nextgroupplot[xlabel={HE-GPT4o}]
      \addplot[fill=academic-blue,   bar shift=-25pt] coordinates {(1,83)};
      \addplot[fill=academic-yellow, bar shift=-15pt] coordinates {(1,89)};
      \addplot[fill=academic-orange, bar shift=-5pt]  coordinates {(1,92)};
      \addplot[fill=academic-purple, bar shift=5pt]   coordinates {(1,93)};
      \addplot[fill=academic-red,    bar shift=15pt]  coordinates {(1,92)};
      \addplot[fill=academic-green,  bar shift=25pt]  coordinates {(1,92)};
      \legend{Naive Entropy, Basic Entropy, FirstN, CoT, Semantic Probability, Semantic Entropy}

    \nextgroupplot[xlabel={HE-GPT3.5-turbo}]
      \addplot[fill=academic-blue,   bar shift=-25pt] coordinates {(1,73)};
      \addplot[fill=academic-yellow, bar shift=-15pt] coordinates {(1,80)};
      \addplot[fill=academic-orange, bar shift=-5pt]  coordinates {(1,84)};
      \addplot[fill=academic-purple, bar shift=5pt]   coordinates {(1,85)};
      \addplot[fill=academic-red,    bar shift=15pt]  coordinates {(1,88)};
      \addplot[fill=academic-green,  bar shift=25pt]  coordinates {(1,89)};

    \nextgroupplot[xlabel={HE-Llama3}]
      \addplot[fill=academic-blue,   bar shift=-25pt] coordinates {(1,72)};
      \addplot[fill=academic-yellow, bar shift=-15pt] coordinates {(1,65)};
      \addplot[fill=academic-orange, bar shift=-5pt]  coordinates {(1,76)};
      \addplot[fill=academic-purple, bar shift=5pt]   coordinates {(1,70)};
      \addplot[fill=academic-red,    bar shift=15pt]  coordinates {(1,76)};
      \addplot[fill=academic-green,  bar shift=25pt]  coordinates {(1,77)};

    \nextgroupplot[xlabel={HE-CodeQwen}]
      \addplot[fill=academic-blue,   bar shift=-25pt] coordinates {(1,86)};
      \addplot[fill=academic-yellow, bar shift=-15pt] coordinates {(1,90)};
      \addplot[fill=academic-orange, bar shift=-5pt]  coordinates {(1,92)};
      \addplot[fill=academic-purple, bar shift=5pt]   coordinates {(1,89)};
      \addplot[fill=academic-red,    bar shift=15pt]  coordinates {(1,93)};
      \addplot[fill=academic-green,  bar shift=25pt]  coordinates {(1,94)};

    \nextgroupplot[xlabel={LeetCode-GPT4o}]
      \addplot[fill=academic-blue,   bar shift=-25pt] coordinates {(1,81)};
      \addplot[fill=academic-yellow, bar shift=-15pt] coordinates {(1,82)};
      \addplot[fill=academic-orange, bar shift=-5pt]  coordinates {(1,87)};
      \addplot[fill=academic-purple, bar shift=5pt]   coordinates {(1,91)};
      \addplot[fill=academic-red,    bar shift=15pt]  coordinates {(1,93)};
      \addplot[fill=academic-green,  bar shift=25pt]  coordinates {(1,94)};

    \nextgroupplot[xlabel={LeetCode-GPT3.5-turbo}]
      \addplot[fill=academic-blue,   bar shift=-25pt] coordinates {(1,77)};
      \addplot[fill=academic-yellow, bar shift=-15pt] coordinates {(1,71)};
      \addplot[fill=academic-orange, bar shift=-5pt]  coordinates {(1,76)};
      \addplot[fill=academic-purple, bar shift=5pt]   coordinates {(1,83)};
      \addplot[fill=academic-red,    bar shift=15pt]  coordinates {(1,90)};
      \addplot[fill=academic-green,  bar shift=25pt]  coordinates {(1,90)};

    \nextgroupplot[xlabel={LeetCode-Llama3}]
      \addplot[fill=academic-blue,   bar shift=-25pt] coordinates {(1,70)};
      \addplot[fill=academic-yellow, bar shift=-15pt] coordinates {(1,59)};
      \addplot[fill=academic-orange, bar shift=-5pt]  coordinates {(1,68)};
      \addplot[fill=academic-purple, bar shift=5pt]   coordinates {(1,70)};
      \addplot[fill=academic-red,    bar shift=15pt]  coordinates {(1,72)};
      \addplot[fill=academic-green,  bar shift=25pt]  coordinates {(1,75)};

    \nextgroupplot[xlabel={LeetCode-CodeQwen}]
      \addplot[fill=academic-blue,   bar shift=-25pt] coordinates {(1,79)};
      \addplot[fill=academic-yellow, bar shift=-15pt] coordinates {(1,77)};
      \addplot[fill=academic-orange, bar shift=-5pt]  coordinates {(1,83)};
      \addplot[fill=academic-purple, bar shift=5pt]   coordinates {(1,82)};
      \addplot[fill=academic-red,    bar shift=15pt]  coordinates {(1,90)};
      \addplot[fill=academic-green,  bar shift=25pt]  coordinates {(1,91)};

    \nextgroupplot[xlabel={MBPP-GPT4o}]
      \addplot[fill=academic-blue,   bar shift=-25pt] coordinates {(1,76)};
      \addplot[fill=academic-yellow, bar shift=-15pt] coordinates {(1,70)};
      \addplot[fill=academic-orange, bar shift=-5pt]  coordinates {(1,72)};
      \addplot[fill=academic-purple, bar shift=5pt]   coordinates {(1,81)};
      \addplot[fill=academic-red,    bar shift=15pt]  coordinates {(1,82)};
      \addplot[fill=academic-green,  bar shift=25pt]  coordinates {(1,83)};

    \nextgroupplot[xlabel={MBPP-GPT3.5-turbo}]
      \addplot[fill=academic-blue,   bar shift=-25pt] coordinates {(1,67)};
      \addplot[fill=academic-yellow, bar shift=-15pt] coordinates {(1,61)};
      \addplot[fill=academic-orange, bar shift=-5pt]  coordinates {(1,63)};
      \addplot[fill=academic-purple, bar shift=5pt]   coordinates {(1,71)};
      \addplot[fill=academic-red,    bar shift=15pt]  coordinates {(1,74)};
      \addplot[fill=academic-green,  bar shift=25pt]  coordinates {(1,74)};

    \nextgroupplot[xlabel={MBPP-Llama3}]
      \addplot[fill=academic-blue,   bar shift=-25pt] coordinates {(1,53)};
      \addplot[fill=academic-yellow, bar shift=-15pt] coordinates {(1,54)};
      \addplot[fill=academic-orange, bar shift=-5pt]  coordinates {(1,54)};
      \addplot[fill=academic-purple, bar shift=5pt]   coordinates {(1,56)};
      \addplot[fill=academic-red,    bar shift=15pt]  coordinates {(1,61)};
      \addplot[fill=academic-green,  bar shift=25pt]  coordinates {(1,61)};

    \nextgroupplot[xlabel={MBPP-CodeQwen}]
      \addplot[fill=academic-blue,   bar shift=-25pt] coordinates {(1,77)};
      \addplot[fill=academic-yellow, bar shift=-15pt] coordinates {(1,70)};
      \addplot[fill=academic-orange, bar shift=-5pt]  coordinates {(1,72)};
      \addplot[fill=academic-purple, bar shift=5pt]   coordinates {(1,76)};
      \addplot[fill=academic-red,    bar shift=15pt]  coordinates {(1,81)};
      \addplot[fill=academic-green,  bar shift=25pt]  coordinates {(1,82)};

    \nextgroupplot[xlabel={TestEval-GPT4o}]
      \addplot[fill=academic-blue,   bar shift=-25pt] coordinates {(1,67)};
      \addplot[fill=academic-yellow, bar shift=-15pt] coordinates {(1,69)};
      \addplot[fill=academic-orange, bar shift=-5pt]  coordinates {(1,71)};
      \addplot[fill=academic-purple, bar shift=5pt]   coordinates {(1,73)};
      \addplot[fill=academic-red,    bar shift=15pt]  coordinates {(1,73)};
      \addplot[fill=academic-green,  bar shift=25pt]  coordinates {(1,74)};
    
    \nextgroupplot[xlabel={TestEval-GPT3.5-turbo}]
      \addplot[fill=academic-blue,   bar shift=-25pt] coordinates {(1,46)};
      \addplot[fill=academic-yellow, bar shift=-15pt] coordinates {(1,45)};
      \addplot[fill=academic-orange, bar shift=-5pt]  coordinates {(1,53)};
      \addplot[fill=academic-purple, bar shift=5pt]   coordinates {(1,57)};
      \addplot[fill=academic-red,    bar shift=15pt]  coordinates {(1,60)};
      \addplot[fill=academic-green,  bar shift=25pt]  coordinates {(1,61)};
    
    \nextgroupplot[xlabel={TestEval-Llama3}]
      \addplot[fill=academic-blue,   bar shift=-25pt] coordinates {(1,50)};
      \addplot[fill=academic-yellow, bar shift=-15pt] coordinates {(1,48)};
      \addplot[fill=academic-orange, bar shift=-5pt]  coordinates {(1,54)};
      \addplot[fill=academic-purple, bar shift=5pt]   coordinates {(1,52)};
      \addplot[fill=academic-red,    bar shift=15pt]  coordinates {(1,58)};
      \addplot[fill=academic-green,  bar shift=25pt]  coordinates {(1,58)};
    
    \nextgroupplot[xlabel={TestEval-CodeQwen}]
      \addplot[fill=academic-blue,   bar shift=-25pt] coordinates {(1,76)};
      \addplot[fill=academic-yellow, bar shift=-15pt] coordinates {(1,83)};
      \addplot[fill=academic-orange, bar shift=-5pt]  coordinates {(1,81)};
      \addplot[fill=academic-purple, bar shift=5pt]   coordinates {(1,83)};
      \addplot[fill=academic-red,    bar shift=15pt]  coordinates {(1,85)};
      \addplot[fill=academic-green,  bar shift=25pt]  coordinates {(1,86)};

    \end{groupplot}

    \node[anchor=south]{\pgfplotslegendfromname{commonlegend}};
  \end{tikzpicture}

  \caption{Validity Rate (VR) comparison across 16 experiments. 
    Existing baselines are 
    \textbf{Naive Entropy} (\textcolor{academic-blue}{\rule[0.5ex]{1.5em}{1.5ex}}), 
    \textbf{Basic Entropy} (\textcolor{academic-yellow}{\rule[0.5ex]{1.5em}{1.5ex}}), 
    \textbf{FirstN} (\textcolor{academic-orange}{\rule[0.5ex]{1.5em}{1.5ex}}), 
    \textbf{Semantic Probability} (\textcolor{academic-red}{\rule[0.5ex]{1.5em}{1.5ex}}), 
    \textbf{Semantic Entropy} (\textcolor{academic-green}{\rule[0.5ex]{1.5em}{1.5ex}}), and 
    \textbf{CoT} (\textcolor{academic-purple}{\rule[0.5ex]{1.5em}{1.5ex}}).}
   \label{baseline_fig}
\end{figure*}

\subsection{\textbf{RQ3: How effective is {\tool} in improving code generation?}}
\label{RQ2}

To assess the impact of {\tool} on code generation, we integrated it into the Reflexion and LATS agents. Since these agents rely on test results to guide their improvement process, the presence of invalid test cases can mislead the model, reducing code correctness. We hypothesize that filtering out invalid test cases using {\tool} will improve Reflexion’s and LATS's effectiveness in generating correct code.


The results, summarized in Table~\ref{table:100}, show that the integration of {\tool} significantly improves code generation performance. On BigCodeBench-hard, Reflexion’s \textit{pass@1} score improved from \textbf{0.30 to 0.37} (a \textbf{7\% increase}), while on BigCodeBench-full, it increased from \textbf{0.56 to 0.67} (an \textbf{11\% increase}). These improvements indicate that {\tool} enhances code correctness by filtering out invalid test cases, thereby improving Reflexion's ability to iteratively refine solutions based on valid feedback.

The results on LATS demonstrate that {\tool}'s benefits extend beyond Reflexion to structurally different agent architectures. On BigCodeBench, integrating {\tool} with LATS improved the pass@1 score from 60\% to 72\%, representing a 12 percentage point improvement. The consistency of improvement across both Reflexion and LATS provides strong evidence that filtering invalid tests with {\tool} is not an artifact of a specific agent design, but rather a generalizable enhancement for any feedback-driven code generation system that relies on LLM-generated test cases for iterative refinement. These results confirm that valid test suites provide more reliable feedback signals regardless of the underlying agent architecture.

The improvement in pass@1 scores when using {\tool} reflects not only better test validity but also enhanced realistic fault detection. In the code generation workflows, the agent generates intermediate code solutions that may contain various types of faults. By filtering out invalid test cases, {\tool} ensures that the feedback provided to the agent is based on tests that can reliably identify these realistic faults, rather than providing misleading signals from hallucinated or logically flawed tests. This enables the agent to more accurately identify and correct defects in the generated code.

\newcommand{\best}[1]{\textbf{#1}}
\subsection{\textbf{RQ4: Comparison with baselines}}
\label{4.2}
In the \textbf{first baseline}, we utilize Naive Entropy instead of semantic entropy. Naive entropy is a fundamental measure of uncertainty, calculated based on the entropy of token sequences rather than focusing exclusively on tokens that contribute to meaning, as is the case with semantic entropy. The primary limitation of Naive Entropy is its inability to recognize that different token sequences may convey the same meaning in natural language. Consequently, when an LLM generates multiple variations of a correct answer, Naive Entropy erroneously inflates entropy values. This baseline follows the same methodology as {\tool}, with the key distinction that, instead of computing features based on the function input and expected output, we extract features from token sequences without selectively considering only the meaningful components.

In the \textbf{second baseline}, we replace Semantic Entropy with Semantic Probability in {\tool}. Instead of relying on token entropy, this approach uses token probability as a measure of uncertainty. A high probability indicates high confidence in the generated token, and a low probability reflects low confidence. Although this baseline is expected to perform similarly to semantic entropy, its performance is likely to be slightly lower, as it considers only the probability of the generated token and does not account for the probability distribution of alternative candidate tokens.

In the third and fourth baselines (FirstN and Basic Semantic Entropy), we use heuristic methods to select the tests. In the FirstN approach, we simply choose the first N generated tests per function and ignore other tests. In the Basic Entropy approach, we calculate the average semantic entropy for each test and select the top N tests with the highest value in average semantic entropy. These baselines don't require any training.

As an additional baseline, we use Chain-of-Thought (CoT) prompting, which validates test cases through step-by-step reasoning based on the prompt and the test case. This method requires a separate LLM call for each test case, making it \textbf{more expensive} in terms of both time and cost. Furthermore, its effectiveness heavily \textbf{depends on the reasoning capability} of the underlying LLM, which may limit its practical use in real-world software projects. In contrast, neither the other baselines nor {\tool} has these limitations.

To evaluate the effectiveness of Semantic Entropy, we compare its performance against five baseline approaches: Naive Entropy, Semantic Probability, FirstN, Basic Entropy, and CoT across 16 different test suites (4 datasets $\times$ 4 LLMs). The results are summarized in Figure~\ref{baseline_fig}, Table~\ref{LC_Bases}, Table~\ref{MS_Bases}, and Table~\ref{tab:method_comparison_ref}.

\subsubsection{Comparison of Test Validity (Precision)}

The results of comparison on VR is presented in Figure~\ref{baseline_fig}. In all experiments, Semantic Entropy outperforms Naive Entropy, FirstN, and Basic Semantic Entropy, by a significant margin. This improvement is expected, as Naive Entropy considers only token sequence entropy, which can misrepresent semantic uncertainty in test case generation. Unlike Naive Entropy, which treats all token variations as equally uncertain, Semantic Entropy selectively evaluates entropy within the context of meaning, leading to \textbf{more accurate filtering of invalid test cases}. The other two baselines are also expected to underperform compared to Semantic Entropy. In the FirstN approach, all tests generated after the $N^{\text{th}}$ test are ignored. In Basic Semantic Entropy, test correctness is determined using only a single statistical feature of the semantic entropy, rather than extracting multiple features and training a classifier.

Semantic Entropy outperforms CoT in 15 experiments, only underperforming on HE-GPT4o by 1\%. While CoT's performance on GPT4o is decent compared to Semantic Entropy, it falls significantly behind on other LLMs that may have less powerful reasoning ability.

When compared to Semantic Probability, Semantic Entropy still shows a \textbf{performance advantage in 11 out of 16 test suites}, while achieving equivalent performance in the remaining five. This result highlights the added value of incorporating entropy-based uncertainty rather than relying solely on token probability. Although Semantic Probability estimates confidence in token generation, it lacks the ability to account for \textbf{alternative plausible token choices} in the model's output distribution. In contrast, Semantic Entropy considers the \textbf{distributional uncertainty} across multiple probable tokens, providing a \textbf{more robust measure of test validity}.

\begin{table}[ht]
\centering
\caption{Wilcoxon Signed-Rank Test Results for Validity Rate (VR): Semantic Entropy vs. Baselines}
\label{tab:wilcoxon_vr}
\setlength{\tabcolsep}{3pt}
\begin{tabular}{@{}lcccc@{}}
\toprule
\textbf{Baseline Method} & \textbf{Statistic} & \textbf{$p$-value} & \textbf{Mean Diff.} & \textbf{Confidence Level} \\
\midrule
Semantic Probability     & 66.0  & 1.24e-3 & 0.0081 & 99.0\%   \\
CoT   & 134.5 & 4.6e-5  & 0.0431 & 99.9\%   \\
FirstN                   & 120.0 & 3.23e-4 & 0.0644 & 99.9\%   \\
Basic Entropy            & 136.0 & 1.5e-5  & 0.1050 & 99.9\%   \\
Naive Entropy            & 136.0 & 1.5e-5  & 0.0925 & 99.9\%   \\
\bottomrule
\end{tabular}
\end{table}

\textbf{Statistical Significance Testing.} 
To substantiate the superiority of VALTEST over baseline methods, we performed Wilcoxon signed-rank tests comparing the VR of Semantic Entropy against each baseline across all 16 experiments. The results demonstrate that Semantic Entropy achieves statistically significantly higher VR values compared to all baseline methods. Specifically, Semantic Entropy significantly outperforms Semantic Probability ($p = 1.24 \times 10^{-3}$), Chain-of-Thought ($p = 4.6 \times 10^{-5}$), FirstN ($p = 3.23 \times 10^{-4}$), Basic Entropy ($p = 1.5 \times 10^{-5}$), and Naive Entropy ($p = 1.5 \times 10^{-5}$), with all comparisons significant at the 99\% to 99.9\% confidence level. Table~\ref{tab:wilcoxon_vr} provides detailed statistical test results for all comparisons.

\begin{table}[t]
\caption{Recall comparison across baseline filtering methods. Values are averaged across all four LLMs for each dataset, with an overall average across datasets. Recall measures the fraction of truly valid tests that are retained after filtering.}
\label{tab:recall_baselines}
\centering
\setlength{\tabcolsep}{3pt}
\renewcommand{\arraystretch}{1}
\begin{tabular}{lcccccc}
\toprule
\textbf{Dataset} & \textbf{Semantic} & \textbf{Semantic} & \textbf{CoT} & \textbf{FirstN} & \textbf{Basic} & \textbf{Naive} \\
 & \textbf{Entropy} & \textbf{Probability} &  &  & \textbf{Entropy} & \textbf{Entropy} \\
\hline
HE       & \best{0.70} & \best{0.70} & 0.67 & 0.41 & 0.38 & \best{0.70} \\
\rowcolor{gray!6}
LeetCode & \best{0.61} & 0.59 & 0.59 & 0.48 & 0.44 & 0.43 \\
MBPP     & 0.40 & 0.41 & 0.40 & \best{0.44} & \best{0.44} & 0.41 \\
\rowcolor{gray!6}
TestEval & \best{0.57} & 0.55 & 0.51 & 0.38 & 0.37 & 0.39 \\
\hline
\textbf{Total AVG} & \best{0.57} & 0.56 & 0.54 & 0.42 & 0.40 & 0.48 \\
\bottomrule
\end{tabular}
\end{table}

\subsubsection{Comparison on Recall}
\label{recal_comp}

\textbf{Semantic Entropy consistently provides the strongest overall performance}, outperforming all baseline approaches in both precision and recall. Semantic Entropy achieves the highest average recall (Table~\ref{tab:recall_baselines}), while also attaining the highest precision (Figure~\ref{baseline_fig}).

The only exception arises with MBPP, where Semantic Entropy shows a \textbf{slight recall deficit} relative to FirstN and Basic Entropy: recall is approximately \textbf{4 percentage points lower}, despite Semantic Entropy maintaining an \textbf{average precision nearly 10 percentage points higher} than both methods. In all other datasets, Semantic Entropy demonstrates superior performance in both recall and precision.

A plausible explanation is that MBPP may be inherently lower in quality or less informative than the other benchmarks. Interestingly, MBPP prompts are substantially shorter than those in others (approximately 90 characters on average, compared to roughly 380 for HumanEval, 690 for TestEval, and 1050 for LeetCode). Shorter prompts may underspecify requirements, increasing the likelihood of ambiguity. This may explain why Semantic Entropy slightly underperforms on the MBPP dataset in terms of recall. Under such conditions, Semantic Entropy's more accurate filtering still yields \emph{substantially better precision}, but this may come at the cost of sacrificing a small amount of recall relative to less accurate approaches that retain more tests indiscriminately. More interestingly, this phenomenon appears limited to lower-quality datasets like MBPP, whereas Semantic Entropy outperforms other methods in both precision and recall on other datasets. Overall, these results indicate that \textbf{Semantic Entropy dominates the baselines in the precision--recall trade-off}.

\subsubsection{Comparison of Test Adequacy}

\begin{table}[h]

\centering
\setlength{\tabcolsep}{3pt}
\renewcommand{\arraystretch}{1}
\caption{Line Coverage (LC) comparison across four datasets. The values represent the average across all four LLMs (GPT-4o, GPT-3.5-turbo, LLama3, and CodeQwen).}
\label{LC_Bases}
\begin{tabular}{lcccccc}
\toprule
\textbf{Dataset} & \textbf{Semantic} & \textbf{Semantic} & \textbf{CoT} & \textbf{FirstN} & \textbf{Basic} & \textbf{Naive} \\
 & \textbf{Entropy} & \textbf{Probability} &  &  & \textbf{Entropy} & \textbf{Entropy} \\
\midrule
HE        & \best{0.86} & 0.85 & 0.83 & 0.82 & 0.80 & 0.81 \\
\rowcolor{gray!6}
LeetCode  & \best{0.96} & 0.95 & 0.94 & 0.94 & 0.93 & 0.92 \\
MBPP      & \best{0.76} & \best{0.76} & 0.74 & 0.73 & 0.70 & 0.72 \\
\rowcolor{gray!6}
TestEval  & \best{0.79} & 0.78 & 0.75 & 0.73 & 0.72 & 0.72 \\
\midrule
\textbf{Total AVG} & \best{0.84} & 0.83 & 0.82 & 0.80 & 0.79 & 0.79 \\
\bottomrule
\end{tabular}
\end{table}

\begin{table}[h]
\centering
\setlength{\tabcolsep}{3pt}
\renewcommand{\arraystretch}{1}
\caption{Mutation Score (MS) comparison across four datasets. The values represent the average across all four LLMs (GPT-4o, GPT-3.5-turbo, LLama3, and CodeQwen).}
\label{MS_Bases}
\begin{tabular}{lcccccc}
\toprule
\textbf{Dataset} & \textbf{Semantic} & \textbf{Semantic} & \textbf{CoT} & \textbf{FirstN} & \textbf{Basic} & \textbf{Naive} \\
 & \textbf{Entropy} & \textbf{Probability} &  &  & \textbf{Entropy} & \textbf{Entropy} \\
\midrule
HE        & \best{0.73} & 0.72 & 0.71 & 0.70 & 0.69 & 0.69 \\
\rowcolor{gray!6}
LeetCode  & \best{0.80} & 0.79 & 0.78 & 0.78 & 0.76 & 0.75 \\
MBPP      & \best{0.72} & \best{0.72} & 0.70 & 0.69 & 0.67 & 0.69 \\
\rowcolor{gray!6}
TestEval  & \best{0.63} & 0.62 & 0.60 & 0.59 & 0.58 & 0.58 \\
\midrule
\textbf{Total AVG} & \best{0.72} & 0.71 & 0.70 & 0.69 & 0.67 & 0.68 \\
\bottomrule
\end{tabular}
\end{table}

To provide a more complete comparison of the baselines, we evaluate their performance using established test adequacy metrics: Line Coverage (LC) and Mutation Score (MS). As shown in Table~\ref{LC_Bases} and Table~\ref{MS_Bases}, these metrics reveal the quality and fault-detection ability of the filtered test suites, which is an important consideration alongside the Validity Rate from Figure~\ref{baseline_fig}.

\begin{table}[ht]
\centering
\footnotesize
\caption{Impact of varying Threshold and topN parameters on {\tool} for GPT-3.5 Turbo on the HumanEval dataset. Metrics include the number of tests generated (Tests), Line Coverage (LC), and Mutation Score (MS). Arrows indicate an increase ({\color{Green}$\uparrow$}) or decrease ({\color{Red}$\downarrow$}) compared to the baseline (Row 1). Cells highlighted in pink indicate parameter changes from the baseline.}
\label{table:2}
\begin{tabular}{
    c
    c
    c
    S[table-format=4.0]
    l
    l
    l
}
\toprule
Row& Threshold & topN & {Tests} & {VR} & {MS} \\
\midrule
1 & 0.75 & 5 & 1553 & 0.892 & 0.79 \\
2 & \cellcolor{pink}0.5 & 5 & 2237 & 0.812 {\color{Red}$\downarrow$ 8.0\%} & 0.821 {\color{Green}$\uparrow$ 3.1\%} \\
3 & \cellcolor{pink}0.65 & 5 & 1958 & 0.843 {\color{Red}$\downarrow$ 4.9\%} & 0.816 {\color{Green}$\uparrow$ 2.6\%} \\
4 & \cellcolor{pink}0.85 & 5 & 1179 & 0.897 {\color{Green}$\uparrow$ 0.5\%} & 0.778 {\color{Red}$\downarrow$ 1.2\%} \\
5 & 0.75 & \cellcolor{pink}1 & 1392 & 0.899 {\color{Green}$\uparrow$ 0.7\%} & 0.769 {\color{Red}$\downarrow$ 2.1\%} \\
6 & 0.75 & \cellcolor{pink}3 & 1415 & 0.898 {\color{Green}$\uparrow$ 0.6\%} & 0.788 {\color{Red}$\downarrow$ 0.2\%} \\
7 & 0.75 & \cellcolor{pink}7 & 1603 & 0.867 {\color{Red}$\downarrow$ 2.5\%} & 0.815 {\color{Green}$\uparrow$ 2.5\%} \\
\bottomrule
\end{tabular}
\end{table}

The results show a clear pattern: baselines that are good at achieving a high VR also produce test suites with better test adequacy scores. This indicates that the methods for identifying valid tests also help in selecting tests that are more effective. Consistent with its superior performance in VR, Semantic Entropy outperforms all other baselines in both LC and MS. Its ability to analyze uncertainty based on meaning, rather than just tokens, allows it to preserve tests that are not only correct but also more valuable for finding faults. This demonstrates that Semantic Entropy provides the best overall strategy for curating a high-quality test suite. Following closely, Semantic Probability also performs well, confirming that methods based on semantic analysis are more effective than simpler approaches. The CoT baseline ranks third, showing competitive results in both LC and MS. The remaining baselines, FirstN, Naive Entropy, and Basic Entropy, rank lower. FirstN relies on the unreliable assumption that earlier generated tests are inherently better. Meanwhile, Basic Entropy and Naive Entropy, which operate at the token level, may discard complex yet correct tests because of their high token-level uncertainty. This limitation accounts for their comparatively weaker performance relative to semantically aware methods.

In conclusion, Semantic Entropy stands out as the most effective approach. It consistently achieves the highest VR while also producing test suites with the best line coverage and mutation scores. This makes it the superior choice for creating test suites that are both valid and powerful at detecting faults.

\subsubsection{Comparison of baselines on Downstream Code Generation}
\label{Secdown}
\begin{table}[ht]
\centering
\caption{Comparison of baselines in code generation task using Reflexion on BigCodeBench Dataset. The numbers show pass@1.}
\label{tab:method_comparison_ref}
\begin{tabular}{lccccc}
\toprule
\textbf{Semantic} & \textbf{Semantic} & \textbf{COT} & \textbf{FirstN} & \textbf{Basic} & \textbf{Naive} \\
\textbf{Entropy} & \textbf{Probability} & & & \textbf{Entropy} & \textbf{Entropy} \\
\midrule
0.675 & 0.668 & 0.648 & 0.635 & 0.628 & 0.608 \\
\bottomrule
\end{tabular}
\end{table}

Table~\ref{tab:method_comparison_ref} reports downstream code-generation accuracy (pass@1) for Reflexion on BigCodeBench when its internal test suites are filtered by different approaches. The proposed Semantic Entropy achieves the best pass@1 of $0.675$, edging Semantic Probability ($0.668$) by $+0.007$, and more clearly surpassing Chain-of-Thought ($0.648$, $+0.027$), FirstN ($0.635$, $+0.040$), Basic Entropy ($0.628$, $+0.047$), and Naive Entropy ($0.608$, $+0.067$). These results demonstrate that Semantic Entropy not only yields the highest test-suite validity upstream, but also translates into the strongest end-task gains for an agentic code generator. These findings indicate that preserving tests identified via semantic-uncertainty cues provides Reflexion with more reliable feedback signals, thereby improving code refinement efficacy on a challenging benchmark.

\subsection{\textbf{RQ5: How does the selection of \textit{threshold} and \textit{topN} hyperparameters impact the balance between MS and VR?}}
\label{app:threshNtopn}

The results, presented in Table \ref{table:2}, highlight the trade-off between the validity rate and the mutation score. The default configuration (row 1) uses a \textit{threshold} of 0.8 and a \textit{topN} of 5. Lowering the \textit{threshold} to 0.5 (row 2) makes the selection less restrictive, decreasing VR by 8\% while increasing MS by 3.1\%. A moderate adjustment to 0.65 (row 3) leads to a 4.9\% drop in VR and a 2.6\% rise in MS. In contrast, raising the \textit{threshold} to 0.85 (row 4) slightly boosts VR by 0.5\% but reduces MS by 0.2\%. These results suggest that a higher \textit{threshold} improves validity but limits mutation score, underscoring the balance between these competing factors.

In rows 5, 6, and 7, we vary the \textit{topN} parameter. Reducing \textit{topN} to 1 (row 5) or 3 (row 6) increases VR but decreases MS relative to the baseline. However, increasing \textit{topN} to 7 (row 7) results in a decrease in VR but an increase in MS. This trend shows that selecting more test cases (higher \textit{topN}) leads to higher MS but lower VR, indicating a trade-off between test quality (VR) and test adequacy (MS) as the number of selected cases increases.

The \textit{threshold} should be adjusted according to the use case. In this experiment, we showed that a higher \textit{threshold} and lower \textit{topN} increase VR but reduce MS. Some may prefer to sacrifice MS to gain VR, while others may prioritize MS. With that in mind, and based on our datasets, we recommend a \textit{threshold} of 0.75 and a \textit{topN} of 3 as default, as it provides a balanced trade-off between VR and MS. A \textit{threshold} of 0.75 is sufficiently restrictive to maintain high VR while sacrificing only a small portion of valid tests. Note that these default values by no means are meant to be the optimal values; thus, to get the best results for new datasets and/or models, one may need to run their hyper-parameter tuning. 


\subsection{\textbf{RQ6: What is the impact of different feature sets in {\tool}?}}
\label{app:feature_impact}

\begin{table}[ht]
\footnotesize
\centering
\caption{The impact of different feature sets on the validity rate of {\tool}. The higher intensity in cell shading indicates more impact.}
\label{table:5}
\begin{tabular}{llcccc}
\toprule
\textbf{Dataset} & \textbf{LLM} & \textbf{Base} & \textbf{FI} & \textbf{EO}& \textbf{FI + EO} \\
\midrule
\multirow{3}{*}{HE}
    & GPT-3.5-Turbo & 0.74 & \cellcolor{cyan!40}0.784 & \cellcolor{cyan!60}0.882 & \cellcolor{cyan!80}0.890 \\
    & GPT-4o        & 0.83 & \cellcolor{cyan!40}0.846 & \cellcolor{cyan!60}0.911 & \cellcolor{cyan!80}0.918 \\
    & Llama3        & 0.63 & \cellcolor{cyan!40}0.647 & \cellcolor{cyan!60}0.701 & \cellcolor{cyan!80}0.769 \\
    & Qwen2.5-Coder  & 0.84 & \cellcolor{cyan!40}0.852 & \cellcolor{cyan!60}0.932 & \cellcolor{cyan!80}0.942 \\
\midrule
\multirow{3}{*}{LeetCode}
    & GPT-3.5-Turbo & 0.63 & \cellcolor{cyan!40}0.844 & \cellcolor{cyan!60}0.881 & \cellcolor{cyan!80}0.895 \\
    & GPT-4o        & 0.75 & \cellcolor{cyan!40}0.865 & \cellcolor{cyan!60}0.922 & \cellcolor{cyan!80}0.937 \\
    & Llama3        & 0.46 & \cellcolor{cyan!40}0.668 & \cellcolor{cyan!60}0.725 & \cellcolor{cyan!80}0.748 \\
    & Qwen2.5-Coder & 0.66 & \cellcolor{cyan!40}0.824 & \cellcolor{cyan!60} 0.897 & \cellcolor{cyan!80}0.907 \\
\midrule
\multirow{3}{*}{MBPP}
    & GPT-3.5-Turbo & 0.60 & \cellcolor{cyan!40}0.603 & \cellcolor{cyan!60}0.715& \cellcolor{cyan!80}0.741 \\
    & GPT-4o        & 0.70 & \cellcolor{cyan!40}0.719 & \cellcolor{cyan!60}0.826 & \cellcolor{cyan!80}0.834 \\
    & Llama3        & 0.53 & \cellcolor{cyan!40}0.55 & \cellcolor{cyan!60}0.597 & \cellcolor{cyan!80}0.610 \\
    & Qwen2.5-Coder  & 0.67 & \cellcolor{cyan!40}0.696 & \cellcolor{cyan!60}0.816 & \cellcolor{cyan!80}0.822 \\
\midrule
\multirow{3}{*}{TestEval}
    & GPT-3.5-Turbo & 0.39 & \cellcolor{cyan!40}0.420 & \cellcolor{cyan!60}0.569 & \cellcolor{cyan!80}0.611 \\
    & GPT-4o        & 0.57 & \cellcolor{cyan!40}0.626 & \cellcolor{cyan!60}0.729 & \cellcolor{cyan!80}0.740 \\
     & Llama3        & 0.38 & \cellcolor{cyan!40}0.448 & \cellcolor{cyan!60}0.542 & \cellcolor{cyan!80}0.585 \\
    & Qwen2.5-Coder  & 0.75 & \cellcolor{cyan!40}0.789 & \cellcolor{cyan!60}0.842 & \cellcolor{cyan!80}0.855 \\
\bottomrule
\end{tabular}%
\end{table}
\begin{table}[ht]
\footnotesize
\centering
\caption{Mann–Whitney U Test p-values between the distributions of valid and invalid tests using the \textit{input\_mean} (FI\_mean) and \textit{output\_mean} (EO\_mean) features. Cells highlighted in green indicate a confidence level of more than 99\% (p-value $< 0.01$).}
\label{table:3}
\begin{adjustbox}{max width=\textwidth}
\begin{tabular}{llcc}
\toprule
\textbf{Dataset} & \textbf{LLM} & \textbf{FI\_mean} & \textbf{EO\_mean} \\
\midrule
\multirow{4}{*}{MBPP} 
 & GPT-4o        & 0.03108 & \cellcolor{green!30}\textbf{2.79e-188} \\
 & GPT-3.5-Turbo & \cellcolor{green!30}\textbf{1.83e-06} & \cellcolor{green!30}\textbf{3.65e-190} \\
 & Llama3        & \cellcolor{green!30}\textbf{2.17e-06} & \cellcolor{green!30}\textbf{5.63e-95} \\
 & Qwen2.5-Coder & \cellcolor{green!30}\textbf{0.00102} & \cellcolor{green!30}\textbf{1.82e-223} \\
\midrule
\multirow{4}{*}{HumanEval}
 & GPT-4o        & 0.81913 & \cellcolor{green!30}\textbf{4.77e-110} \\
 & GPT-3.5-Turbo & \cellcolor{green!30}\textbf{0.00444} & \cellcolor{green!30}\textbf{3.56e-131} \\
 & Llama3        & 0.02981 & \cellcolor{green!30}\textbf{1.19e-29} \\
 & Qwen2.5-Coder       & \cellcolor{green!30}\textbf{0.00083} & \cellcolor{green!30}\textbf{4.64e-153} \\
\midrule
\multirow{4}{*}{LeetCode}
 & GPT-4o        & \cellcolor{green!30}\textbf{4.57e-23} & \cellcolor{green!30}\textbf{0.0} \\
 & GPT-3.5-Turbo & \cellcolor{green!30}\textbf{8.63e-36} & \cellcolor{green!30}\textbf{0.0} \\
 & Llama3        & 0.01996 & \cellcolor{green!30}\textbf{1.65e-32} \\
  & Qwen2.5-Coder & \cellcolor{green!30}\textbf{4.65e-05} & \cellcolor{green!30}\textbf{0.0} \\
\midrule
\multirow{4}{*}{TestEval}
& GPT-4o        & \cellcolor{green!30}\textbf{9.20e-05} & \cellcolor{green!30}\textbf{1.07e-127}\\
& GPT-3.5-Turbo & \cellcolor{green!30}\textbf{0.00318} & \cellcolor{green!30}\textbf{2.08e-38} \\
& Llama3        & \cellcolor{green!30}\textbf{0.00231} & \cellcolor{green!30}\textbf{9.23e-65} \\
& Qwen2.5-Coder & \cellcolor{green!30}\textbf{1.15e-07} & \cellcolor{green!30}\textbf{3.12e-56} \\
\bottomrule

\end{tabular}
\end{adjustbox}
\end{table}

\begin{table}[ht]
\footnotesize
\centering
\caption{The average of feature differences between valid and invalid test cases. Higher intensity in cell shading indicates greater difference. }
\label{table:4}
\begin{adjustbox}{max width=0.9\textwidth}
\begin{tabular}{lcc}
\toprule
\textbf{Dataset/LLM} & \(\Delta\)FI\_mean & \(\Delta\)EO\_mean \\
\midrule
MBPP/GPT-4o            & \cellcolor{green!20}0.03  & \cellcolor{green!40}0.23 \\
MBPP/GPT-3.5-Turbo      & \cellcolor{red!20}$-$0.06 & \cellcolor{green!40}0.30 \\
MBPP/Llama3            & \cellcolor{red!20}$-$0.07 & \cellcolor{green!40}0.27 \\
MBPP/Qwen2.5-Coder     & \cellcolor{red!20}$-$0.07 & \cellcolor{green!40}0.27 \\
\midrule
HumanEval/GPT-4o       & \cellcolor{red!10}$-$0.01 & \cellcolor{green!50}0.41 \\
HumanEval/GPT-3.5-Turbo & \cellcolor{green!20}0.05  & \cellcolor{green!50}0.40 \\
HumanEval/Llama3       & \cellcolor{red!20}$-$0.07 & \cellcolor{green!30}0.15 \\
HumanEval/Qwen2.5-Coder  & \cellcolor{green!20}0.04 & \cellcolor{green!40}0.21 \\
\midrule
LeetCode/GPT-4o        & \cellcolor{green!40}0.11  & \cellcolor{green!80}0.64 \\
LeetCode/GPT-3.5-Turbo  & \cellcolor{green!40}0.11  & \cellcolor{green!70}0.57 \\
LeetCode/Llama3        & \cellcolor{red!10}$-$0.02 & \cellcolor{green!30}0.15 \\
LeetCode/Qwen2.5-Coder  & \cellcolor{red!10}$-$0.01 & \cellcolor{green!40}0.22 \\
\midrule
TestEval/GPT-4o & \cellcolor{red!20}$-$0.07 & \cellcolor{green!50}0.45 \\
TestEval/GPT-3.5-Turbo & \cellcolor{red!20}$-$0.07 & \cellcolor{green!40}0.29 \\
TestEval/Llama3 & \cellcolor{red!10}$-$0.01 & \cellcolor{green!40}0.18 \\
TestEval/Qwen2.5-Coder & \cellcolor{green!20}0.06 & \cellcolor{green!40}0.21\\

\bottomrule

\end{tabular}
\end{adjustbox}
\end{table}

For this ablation study, we first analyze whether there are significant differences in entropy values between valid and invalid test cases. The hypothesis is that when LLM is less confident in the generated tokens, the test is more likely to be invalid; thus, the token entropy values are higher in invalid tests, indicating hallucination. As explained in the design, we select two features (\textit{FI\_mean} and \textit{EO\_mean}), representing the mean token entropy values from the two feature sets. Then, we apply the Mann-Whitney U test to compare the distributions of these features between valid and invalid test cases. This test helps determine whether there is a significant difference between each pair of distributions (in valid vs. invalid test cases), per feature. The results are presented in Table \ref{table:3}. We conducted this test on the two features across the 16 generated test suites. Cells with a p-value less than 0.01 are colored green. Typically, a p-value of less than 0.05 indicates a statistically significant difference between the two distributions.

\begin{table*}[ht!]
\centering
\footnotesize
\caption{Comparison of test generation metrics for GPT-4o across datasets under VALTEST and VALTEST with Correction settings. Metrics include the number of tests generated (Tests), Validity Rate (VR), Line Coverage (LC), and Mutation Score (MS). Percentage changes are indicated with colored arrows: decreases in red and increases in green.}
\label{table:6}
\begin{tabular}{
    ll
    S[table-format=4.0]
    S[table-format=1.3]
    S[table-format=1.3]
    S[table-format=1.2]
    S[table-format=4.0]
    l
    S[table-format=1.3]
    l
}
\toprule
\multicolumn{2}{c}{} & \multicolumn{4}{c}{VALTEST} & \multicolumn{4}{c}{VALTEST + Correction} \\
\cmidrule(lr){3-6} \cmidrule(lr){7-10}
Dataset & LLM & {\#Tests} & {VR} & {LC} & {MS} & {\#Tests} & {VR} & {LC} & {MS} \\
\midrule
HE & GPT-4o & 2512 & 0.92 & 0.96 & 0.86 & 3157 & 0.91\,{\color{Red}($-1$\% $\downarrow$)} & 0.97 & 0.87\,{\color{Green}($+1$\% $\uparrow$)} \\
LeetCode & GPT-4o & 5651 & 0.94 & 0.98 & 0.85 & 9047 & 0.85\,{\color{Red}($-9$\% $\downarrow$)} & 0.99 & 0.88\,{\color{Green}($+3$\% $\uparrow$)} \\
MBPP & GPT-4o & 3581 & 0.83 & 0.81 & 0.78 & 7750 & 0.78\,{\color{Red}($-5$\% $\downarrow$)} & 0.87 & 0.86\,{\color{Green}($+8$\% $\uparrow$)} \\
\bottomrule
\end{tabular}
\end{table*}

The feature set \textit{EO\_mean} shows significant differences between valid and invalid test cases in all test suites. This suggests that the token entropy value in the expected output section of a test case differs significantly between valid and invalid cases. In contrast, the feature set \textit{FI\_mean} shows notable differences between valid and invalid test cases in some cases but not all. These observations suggest that LLMs exhibit different patterns when generating valid versus invalid test cases, particularly in the token entropy values of the function input and expected output sections of the tests. The difference is more pronounced in the expected output tokens, where LLMs tend to ``hallucinate'' when uncertain, often producing invalid output.

To further explore this finding, in Table \ref{table:4}, we present the differences in the average values of the two features between valid and invalid test cases. For instance, \textit{$\Delta$EO\_mean} represents the difference between the average of \textit{EO\_mean} for invalid test cases and the average of \textit{EO\_mean} for valid test cases, i.e., $\textit{AVG(EO\_mean}_{\text{invalid}}) - \textit{AVG(EO\_mean}_{\text{valid}})$. This table demonstrates that invalid test cases have higher entropy values in \textit{EO\_mean} compared to valid ones. This suggests that when LLMs are less certain about the generated output tokens (high entropy), they tend to hallucinate, which results in invalid test cases. However, these experiments did not fully reveal the actual impact of the feature sets on the performance of {\tool}.

In the final experiments for this research question, we conducted a proper ablation study to assess the impact of different feature sets, \textit{EO} and \textit{FI}, on the performance of {\tool}. The results, presented in Table \ref{table:5}, report the VR for each experiment. Specifically, we evaluated the performance using \textit{FI} only, \textit{EO} only, and a combination of both feature sets. The findings indicate that \textit{EO} feature set consistently has a greater influence on the validity rate of {\tool} compared to \textit{FI} feature set. Moreover, incorporating both feature sets yields better results than using \textit{EO} or \textit{FI} alone, demonstrating that all feature sets contribute to the overall performance of {\tool}.

\subsection{\textbf{RQ7: How to combine {\tool} and Chain-of-Thought (CoT) reasoning to improve both validity rate and mutation score of test suites?}}
\label{curation}
The results are presented in Table \ref{table:6}. We compared these findings with the results of {\tool} in the main experiment of the paper. In this experiment, we corrected the invalid test cases identified in \ref{RQ1} and added the corrected cases to the already valid ones to form a new test suite. The number of test cases in these experiments matches the number of tests in the ``Base'' experiments from \ref{RQ1}, as no tests were discarded; instead, they were corrected. A CoT reasoning approach was used to correct the identified invalid test cases. The results demonstrate a substantial improvement in MS compared to {\tool}. Although VR decreases compared to {\tool}, they still improve relative to the ``Base'', in \ref{RQ1}.
MS shows an increase of 1\% to 8\% compared to {\tool} in \ref{RQ1}, while VR decreases by 1\% to 9\%. These results suggest that combining {\tool} with CoT prompting to correct test cases produces a richer test suite that enhances MS without significantly compromising VR in {\tool}, which makes {\tool}+Correction better than Base both in VR and MS.

\begin{table*}[ht]
\footnotesize
\centering
\caption{Categorization of invalid test cases generated by GPT4o on HumanEval. The accuracy column shows the accuracy of {\tool} in detecting invalid tests.}
\label{table:7}
\small
\begin{adjustbox}{max width=\textwidth}
\begin{tabular}{llrrrrr}
\toprule
\multirow{2}{*}{Row} & \multirow{2}{*}{Category} & \multicolumn{2}{c}{Correct Prediction} & \multicolumn{2}{c}{Incorrect Prediction} & \multirow{2}{*}{Accuracy (\%)} \\
\cmidrule(r){3-4} \cmidrule(r){5-6}
& & Count & Percentage (\%) & Count & Percentage (\%) &  \\
\midrule
1 & Misunderstanding of Function Logic & 213 & 61.7 & 98 & 55.4 & 68.5 \\
2 & Off-by-One or Range Errors & 45 & 13.0 & 22 & 12.4 & 67.2 \\
3 & Ambiguities or Incompleteness in the Docstring & 48 & 13.9 & 30 & 16.9 & 61.5 \\
4 & Incorrect Input-Output Mapping & 23 & 6.7 & 5 & 2.8 & 82.1 \\
5 & Edge Case Misinterpretation & 10 & 2.9 & 4 & 2.3 & 71.4 \\
6 & Rounding Expectations & 6 & 1.7 & 10 & 5.6 & 37.5 \\
7 & Others & 0 & 0 & 8 & 4.5 & 0.0 \\
\midrule
\textbf{Total} & & 345 & 100.0 & 177 & 100.0 & 66.1 \\
\bottomrule
\end{tabular}
\end{adjustbox}
\end{table*}
\section{Causal Analysis of LLM-Generated Test Failures and {\tool}’s Detection Efficacy}
\label{app:causal_analysis}

To go deeper into the results and understand when {\tool} is effective and when it is not, in this section, we analyze the types of invalid test cases generated by LLMs. 
Specifically, we categorize invalid test cases based on the root cause of their failure and identify which categories are detectable by {\tool} and which are not. 
We use the HE-GPT4o test suite as a case study for this section. To define these categories, we initially conducted a manual analysis of 30 random test cases (out of 522 total invalid tests) to create a preliminary categorization scheme. Subsequently, we selected another 30 random test cases and categorized them using OpenAI’s o1-preview model, either fitting them into our initial categories or expanding the scheme as necessary. Following a manual verification (sanity check) of the 30 samples' labels, we ended up with six categories that encompass all primary causes of failure for tests generated by GPT4o on the HumanEval dataset. We then tasked the o1-preview model with classifying all test cases into these defined categories, instructing it to select the closest match for cases that could reasonably fit multiple categories. The prompt provided to the model includes the function docstring, code implementation, and the specific failed test cases. 

The six categories are: ``Misunderstanding of Function Logic'', ``Off-by-One or Range Errors'', ``Ambiguities or Incompleteness in the Docstring'', ``Incorrect Input-Output Mapping'', ``Edge Case Misinterpretation'', and ``Rounding Expectations''. The ``Misunderstanding of Function Logic'' category (Category 1) includes cases where the LLM overlooks aspects of the function docstring, generating tests that contradict the specifications, leading to invalid tests. This category contains the highest number of cases overall. Categories 2 and 3 follow the first category in terms of frequency. ``Off-by-one or Range Errors'' (Category 2) occur when the test case expects results that are slightly off due to an off-by-one or range-related misunderstanding. ``Ambiguities or Incompleteness in the Docstring'' (Category 3) arises when the docstring is ambiguous, incomplete, or lacks details, causing the LLM to generate test cases based on incorrect assumptions, leading to assertion errors when run on the function implementation. Although the final LLM-generated categories might not be perfect, looking at the accuracy of {\tool} per category still gives us an understanding of the underlying reasons why {\tool} is not effective.
Table~\ref{table:7} presents the categorization results and the accuracy of {\tool} in detecting invalid tests per category. Category 4 has the best accuracy on {\tool}, while Category 3, where ambiguities in the docstring lead the LLM to make incorrect assumptions, has the lowest accuracy after Category 6. This finding indicates that ambiguous docstrings impair {\tool}’s detection effectiveness more than categories where the docstring is clear. For example, consider the function \textit{vowels\_count}:

\begin{lstlisting}[basicstyle=\footnotesize]
def vowels_count(s):
    """Write a function vowels_count which takes a string representing a word as input and returns the number of vowels in the string. Vowels in this case are 'a', 'e', 'i', 'o', 'u'. Here, 'y' is also a vowel, but only when it is at the end of the given word."""
    vowels = "aeiouAEIOU"
    n_vowels = sum(c in vowels for c in s)
    if s[-1] == 'y' or s[-1] == 'Y':
        n_vowels += 1
    return n_vowels
\end{lstlisting}

And the generated tests:

\begin{lstlisting}[basicstyle=\footnotesize]
Test Case1: assert vowels_count("") == 0
Test Case2: assert vowels_count("crypt") == 1
\end{lstlisting}

Both tests fail when applied to the \textit{vowels\_count} implementation. Note that the tests are generated using only the docstring and function signature; the LLM has not accessed the function’s implementation during test generation. Test case 1 is categorized under Category 3. Here, the LLM assumes that an empty string should yield 0, as there are no vowels. However, in the function implementation, an empty string input raises an exception at line 5. Since this behavior is not described in the docstring, the test falls into Category 3 due to the ambiguity. In test case 2, the LLM incorrectly treats the letter \textit{y} in \textit{crypt} as a vowel, contradicting the explicit instructions in the docstring, placing it in Category 1 due to a misunderstanding of function logic.

{\tool} does not detect the invalidity of the first test case because the generated test case does not conflict with the function docstring, providing no indication of hallucination. However, {\tool} correctly identifies the second test case as invalid since it contradicts the docstring, resulting in an indication of hallucination. This example illustrates that {\tool} effectively detects invalid tests associated with hallucination indicators but struggles with cases that lack such indicators.
\section{Test Semantic Extraction Examples}
\label{Semantic_test}
{\tool} supports Python class-based unit tests in addition to simple one-line assertions. Class-based tests offer greater flexibility for testing in Python, allowing developers to write more comprehensive and structured test cases. In this section, we present examples of identifying the function input and expected output components in Python unit tests, illustrating how Algorithm~\ref{alg:semantic-extraction-try-catch} operates in class-based unit tests.

In the following test case:

\begin{lstlisting}[basicstyle=\footnotesize]
class TestMatchWhatAndLike(unittest.TestCase):
    def setUp(self):
        self.df = pd.DataFrame({
            "Title": ["What is this", "Something you may Like"],
            "Content": ["What do you think?", "I like what you did!"]
        })
    def test_extract_expected_words(self):
        output = {"What": 1, "do": 1, "you": 2,
                  "think": 1, "I": 1, "like": 1,
                  "what": 1, "did": 1}
        assert task_func(self.df) == output
\end{lstlisting}

The algorithm locates the assertion statement \lstinline!assert task_func(self.df) == output! and extracts the function input variable \lstinline!self.df! as well as the expected output variable \lstinline!output!. It then resolves these variables to their concrete values by tracing their definitions within the test case. The resolved function input is:
\begin{lstlisting}[basicstyle=\footnotesize]
pd.DataFrame({
        "Title": ["What is this", "Something you may Like"],
        "Content": ["What do you think?", "I like what you did!"]
})
\end{lstlisting}
The expected output is determined as:
\begin{lstlisting}[basicstyle=\footnotesize]
{"What": 1, "do": 1, "you": 2,
"think": 1, "I": 1, "like": 1,
"what": 1, "did": 1}
\end{lstlisting}
In the following test case:

\begin{lstlisting}[basicstyle=\footnotesize]
class TestListOfNumbers(unittest.TestCase):
    def setUp(self):
        self.input_json = "{'a': [1, 2, 3]}"

    def test_list_elements_doubled(self):
        df = task_func(self.input_json)
        self.assertListEqual(df["a"].tolist(), [2, 4, 6])
\end{lstlisting}

The algorithm identifies the assertion statement at line 7. However, since it does not find the function input on the left-hand side and the expected output on the right-hand side of the assertion, it uses a heuristic method. This method extracts values defined in the \texttt{setUp} method of the class as the function input. It then considers the assertion statement and any values defined within the test method, except those used directly in the assertion, as the expected output. Using this approach, the value \lstinline!"{'a': [1, 2, 3]}"! is extracted as the function input, and the expression \lstinline!self.assertListEqual(df["a"].tolist(), [2, 4, 6])! is identified as the expected output.

The same method is applied to the following test case, where \lstinline!"{}"! is extracted as the function input and \lstinline!self.assertTrue(df.empty)! is identified as the expected output.

\begin{lstlisting}[basicstyle=\footnotesize]
class TestEmptyJSON(unittest.TestCase):
    def setUp(self):
        self.input_json = "{}"

    def test_returns_empty_df(self):
        df = task_func(self.input_json)
        self.assertTrue(df.empty)
\end{lstlisting}

For the following test case, the algorithm extracts \lstinline!"name,age\nAlice,30\nBob,25\n"! as the function input and identifies \lstinline!self.assertTrue(all(df["Status"] == " "))! as the expected output.

\begin{lstlisting}[basicstyle=\footnotesize]
class TestIdenticalFiles(unittest.TestCase):
    def setUp(self):
        self.file1 = tempfile.NamedTemporaryFile()
        self.file2 = tempfile.NamedTemporaryFile()
        content = "name,age\nAlice,30\nBob,25\n"
        self.file1.write(content)
        self.file2.write(content)
        self.file1.close()
        self.file2.close()

    def tearDown(self):
        os.remove(self.file1.name)
        os.remove(self.file2.name)

    def test_no_difference(self):
        df = task_func(self.file1.name, self.file2.name)
        self.assertTrue(all(df["Status"] == " "))
\end{lstlisting}

{\tool} supports test cases that involve exception handling. For instance, in the following test case, the algorithm correctly identifies \lstinline!self.assertRaises(ValueError)! as the expected output, and extracts $10$ and $0$ as the function input. In this case, the \lstinline!self.assertRaises(ValueError)! statement specifies the expected behavior when the function is invoked with the input arguments $10$ and $0$.

\begin{lstlisting}[basicstyle=\footnotesize]
import unittest

class TestDivide(unittest.TestCase):

    def setUp(self):
        self.a = 10
        self.b = 0 

    def test_divide_by_zero(self):
        """Test that dividing by zero raises ValueError."""
        with self.assertRaises(ValueError) as context:
            divide(self.a, self.b)
\end{lstlisting}

\section{Algorithms}
\label{sec:algorithms}
\subsection{Test Semantic Extraction}

\begin{algorithm}[ht]
\footnotesize
\DontPrintSemicolon
\SetAlgoLined
\SetKwInOut{Input}{Input}
\SetKwInOut{Output}{Output}
\Input{Test case $t_i$ as source code}
\Output{$\mathit{input\_str}$, $\mathit{output\_str}$}
\BlankLine
$\mathit{AST} \gets \mathsf{ParseAST}(t_i)$\;
$\mathit{assert\_stmt} \gets \mathsf{AssertFind}(\mathit{AST})$\;
\BlankLine
\Try{
    $\mathit{call} \gets \mathit{assert\_stmt.test.left}$\;
    $\mathit{expected} \gets \mathit{assert\_stmt.test.comparators[0]}$\;
    \If{$\mathsf{is\_variable}(\mathit{call})$}{
        $\mathit{call} \gets \mathsf{FindVariableDefinition}(\mathit{AST}, \mathit{call})$\;
    }
    \If{$\mathsf{is\_variable}(\mathit{expected})$}{
        $\mathit{expected} \gets \mathsf{FindVariableDefinition}(\mathit{AST}, \mathit{expected})$\;
    }
    $\mathit{input\_str} \gets \mathsf{ast.unparse}(\mathit{call}).\mathsf{strip}()$\;
    $\mathit{output\_str} \gets \mathsf{ast.unparse}(\mathit{expected}).\mathsf{strip}()$\;
}
\Catch{ \tcp{Fallback heuristic block}
    $\mathit{input\_str} \gets \mathsf{ExtractValuesFromSetUp}(\mathit{AST})$\;
    $\mathit{output\_str1} \gets \mathsf{ExtractValuesFromTestMethod}(\mathit{AST})$\;
    $\mathit{output\_str} \gets \mathit{output\_str1} \,\texttt{.concat(}\mathit{assert\_stmt}\texttt{)}$
}
\Return{\(\mathit{input\_str}, \mathit{output\_str}\)}\;
\caption{Test Semantic Extraction}
\label{alg:semantic-extraction-try-catch}
\end{algorithm}

Algorithm~\ref{alg:semantic-extraction-try-catch} describes the procedure for extracting the function input and expected output from a test case. The algorithm first constructs an abstract syntax tree (AST) from the source code of the test case, which provides a structured representation suitable for programmatic analysis. It then identifies the assertion statement and attempts to extract the function input from the left-hand side of the assertion and the expected output of the assertion expression, respectively. If either the function input or the expected output is represented as a variable, the algorithm traverses the AST to resolve the variable definition and retrieve the corresponding concrete value. The resolved nodes are subsequently converted into their textual representations using the AST unparsing method.

If this extraction process fails, the algorithm applies a fallback heuristic. This heuristic estimates the function input by analyzing variable assignments in the setup method of the test class to find every concrete value in the setup method. It also approximates the expected output by combining values extracted from variable definitions in the test function and the content of the assertion statement. This design increases the robustness of the extraction process in cases where the test case is more complex.

\subsection{Greedy Token Matching}
\begin{algorithm}[ht]
\footnotesize
\caption{Greedy Token Matching}
\label{matching}
\footnotesize
\DontPrintSemicolon
\SetAlgoLined
\SetKwInOut{Input}{Input}\SetKwInOut{Output}{Output}

\Input{Function input string \(s_{in}\), Expected output string \(s_{out}\), Token iterator \(\mathit{tokenItr}\) over \(\mathbf{W}_i\)}
\Output{Matched tokens \(\mathbf{W}_{in}\) and \(\mathbf{W}_{out}\)}
\BlankLine

\SetKwFunction{MatchTokens}{MatchTokens}
\SetKwProg{Fn}{Function}{:}{}
\Fn{\MatchTokens{\(s\)}}{
    \(\mathit{remaining} \gets \texttt{clean}(s)\)\;
    \(\mathbf{W}_s \gets \emptyset\)\;
    \(\mathit{index} \gets 0\)\;
    \While{\( \mathit{tokenItr}\) is not exhausted}{
        \(temp \gets \texttt{next}(\mathit{tokenItr})\)\;
        \If{\(\mathit{index} = 0\)}{
            \(temp \gets \texttt{lstrip}(temp)\)\;
        }
        \If{\(\mathit{remaining}[\mathit{index}:]\) \texttt{ starts with } \(temp\)}{
            \(\mathbf{W}_s \gets \mathbf{W}_s \cup \{temp\}\)\;
            \(\mathit{index} \gets \mathit{index} + |\texttt{temp}|\)\;
        }
        \If{\(\mathit{index} \ge |\mathit{remaining}|\)}{
            \textbf{break}\;
        }
    }
    \(\mathbf{W}_s \gets \texttt{remove\_unnecessary\_tokens}(\mathbf{W}_s)\)\;
    \Return{\(\mathbf{W}_s\)}\;
}

\(\mathbf{W}_{in} \gets \MatchTokens(s_{in})\)\;
\(\mathbf{W}_{out} \gets \MatchTokens(s_{out})\)\;

\Return{\(\mathbf{W}_{in}, \mathbf{W}_{out}\)}\;

\end{algorithm}

Algorithm~\ref{matching} leverages a greedy token matching strategy to match substrings from function input and expected output strings with tokens drawn from a shared token sequence $\mathbf{W}_i$. Greedy token matching is commonly used in lexical analysis and tokenization processes~\cite{alfred2007compilers}. First, the process matches the input string against tokens provided by an iterator over $\mathbf{W}_i$, in which each token is matched with the starting segment of the remaining string and, after a successful match, is added to the input token set and the matching index advances by the token’s length. Matching proceeds until the input string is fully consumed, after which the same procedure is applied to the expected output string using the residual tokens from $\mathbf{W}_i$. A post-processing step then removes any punctuation tokens from the resulting sets. 

\section{Limitations}
\label{threats}
\textbf{False Negatives.}  
Some test cases that should have been identified as positives may be incorrectly classified as negatives because they are not recognizable by {\tool}. However, our analysis showed that {\tool} has minimal impact on the Mutation Score (MS) and Line Coverage (LC) in most cases, indicating that it retains many valuable test cases that contribute to maintaining high test quality scores. Additionally, we examined the trade-off between test validity and mutation scores in Section~\ref{app:threshNtopn}, demonstrating that {\tool} can be fine-tuned to balance these factors based on specific needs.


\textbf{False Positives.}  
While {\tool} retains valid test cases, there is a risk of mistakenly preserving invalid ones. However, our analysis indicates that this issue is minimal. {\tool} achieves on average a high Validity Rate (VR) of 88\%, 87\%, and 75\%, demonstrating its strong performance in retaining valid tests while effectively discarding invalid ones.

\textbf{Language-Specific Dataset Limitation.} The five datasets used in this study are specific to the Python programming language. Consequently, the results may not be fully generalized to other languages such as Java, JavaScript, or PHP. However, given the nature of LLMs and the language-independent design of {\tool}, the findings could be extended to other programming languages with minimal adaptation. The underlying principles of {\tool} are designed to be language-agnostic. The core concept of our framework, input arguments and expected output, is a fundamental aspect of unit testing across virtually all programming paradigms and languages (e.g., Java with JUnit, Rust with \texttt{assert\_eq!}, C++ with GTest). The primary language-specific component in our pipeline is the test case parser, which is responsible for extracting these semantic elements from the Abstract Syntax Tree (AST). Therefore, adapting {\tool} to a new language would primarily require implementing a new parser tailored to that language's syntax and its standard testing frameworks. The subsequent steps of the pipeline, including entropy calculation and the machine learning-based classification, would remain largely unchanged.

\textbf{Overfitting Risk.} The \textit{threshold} used for filtering test cases based on model predictions may lead to overfitting to specific datasets, potentially introducing bias in the results on these benchmarks. To mitigate this, the \textit{threshold} should be adjusted based on the specific use case to ensure broader applicability. Additionally, to reduce the risk of results being inadvertently biased by specific train-test splits within the benchmark, we employed cross-dataset validation and demonstrated that {\tool} is transferable across datasets, indicating that it does not overfit to any dataset.

\textbf{Reliance on Intrinsic Test Case Signals.} {\tool} derives its predictive signals from the intrinsic syntactic and semantic structure of the generated test cases themselves, without incorporating other sources of information such as natural language problem descriptions, formal specifications, or requirement documents. We argue that this limitation does not significantly undermine the effectiveness of our approach for several reasons. First, when an LLM generates a test case based on a problem description, it implicitly encodes the relevant requirements and constraints into the structure and semantics of the test itself. Therefore, the test case serves as a distilled representation of the LLM's understanding of the problem. Second, our empirical results demonstrate that {\tool} achieves substantial improvements in validity rate and downstream code generation performance without requiring external signals, suggesting that the test cases themselves contain sufficient information for effective validation. Third, relying solely on intrinsic signals ensures that {\tool} remains broadly applicable across diverse domains and scenarios where external documentation may be unavailable, incomplete, or inconsistent. Nevertheless, we acknowledge that incorporating external signals represents a promising direction for future work. Techniques such as Retrieval-Augmented Generation (RAG) could be employed to ground the validation process in requirement documents, potentially leading to more context-aware assessments in scenarios where high-quality external documentation is available.

\textbf{topN Parameter Design Choice.}
The \texttt{topN} parameter in {\tool} was primarily designed to facilitate consistent evaluation on different validation approaches rather than to reflect an optimal development strategy. By ensuring that each function retains at least $N$ test cases (even when all fall below the validity threshold), we enable fair comparison using metrics like Mutation Score and Line Coverage, which would otherwise become undefined for empty test suites. However, this design choice introduces a tension with our core motivation that invalid tests provide harmful feedback. We acknowledge that in production environments, \textbf{providing no feedback when all tests fall below the threshold may be preferable to providing low-confidence tests}. Our evaluation did not empirically assess this alternative strategy. Future work should investigate the comparative effectiveness of strict thresholding (no \texttt{topN}, allowing empty test suites) versus flexible thresholding (\texttt{topN} $>$ 0) in iterative code generation workflows.

\section{Conclusion and Future Work}

In this paper, we introduced \textbf{{\tool}}, a novel framework that employs Semantic Entropy to validate LLM-generated test cases without requiring code execution. Our evaluation across multiple datasets and LLMs demonstrated that \textbf{{\tool} improves test validity rates by up to 29\%}, making it a reliable solution for filtering and refining LLM-generated test cases. Furthermore, by integrating \textbf{{\tool}} into an LLM-based code generation pipeline, we showed that eliminating invalid test cases significantly improves \textit{pass@1} by up to \textbf{11\%}, leading to more accurate and effective automated coding solutions.
Despite its success, {\tool} has areas for improvement. Future work can focus on \textbf{extending {\tool} to other programming languages} and \textbf{implementing an iterative approach} to make test improvement more effective.

Additionally, future work should include the development and evaluation of {\tool} on explicit, realistic fault detection benchmarks. This would involve creating datasets of LLM-generated faulty code with labeled fault types and systematically evaluating the fault detection effectiveness of {\tool}. Such benchmarks would provide more direct and fine-grained evidence of {\tool}'s ability to improve test suites for realistic fault detection in LLM-based code generation scenarios.


\bibliographystyle{IEEEtran}
\bibliography{main}
\newpage

 





\end{document}